\documentclass{aastex6}

\usepackage{graphicx}
\usepackage{textcomp}
\usepackage{natbib}
\usepackage{subfigure}
\usepackage{amsmath}
\usepackage{amssymb}
\usepackage{multirow}\usepackage[section] {placeins}
\begin{document}

\title{The Equation of State of MH-III: A Possible Deep CH$_4$ Reservoir in Titan, Super-Titan Exoplanets and Moons}

\author{A. Levi}
\affiliation{Harvard-Smithsonian Center for Astrophysics, 60 Garden Street, Cambridge, MA 02138, USA}
\email{amitlevi.planetphys@gmail.com}

\author{R. E. Cohen}
\affil{Extreme Materials Initiative, Geophysical Laboratory, Carnegie Institution for Science 5251 Broad Branch Rd. NW, Washington, DC 20015, USA}
\affil{Department of Earth and Environmental Sciences, Ludwig-Maximilians University Munich, Theresienstrasse 41, Munich 80333, Germany}

\maketitle

\section*{ABSTRACT}

We investigate the thermal equation of state, bulk modulus, thermal expansion coefficient, and heat capacity of MH-III (CH$_4$ filled-ice Ih), needed for the study of CH$_4$ transport and outgassing for the case of Titan and super-Titans.  
We employ density functional theory and \textit{ab initio} molecular dynamics simulations in the generalized-gradient approximation  with a van der Waals functional. We examine the finite temperature range of $300$\,K-$500$\,K and pressures between $2$\,GPa-$7$\,GPa. We find that in this P-T range MH-III is less dense than liquid water.

There is uncertainty in the normalized moment of inertia (MOI) of Titan; it is estimated to be in the range of $0.33-0.34$. 
If Titan's MOI is $0.34$, MH-III is not stable at present in Titan's interior, yielding an easier path for the outgassing of CH$_4$. However, for an MOI of $0.33$, MH-III is thermodynamically stable at the bottom of a ice-rock internal layer capable of storing CH$_4$. For rock mass fractions $\lessapprox 0.2$ upwelling melt is likely hot enough to dissociate MH-III along its path. For super-Titans considering a mixture of MH-III and ice VII, melt is always positively buoyant if the H$_2$O:CH$_4$ mole fraction is $>5.5$.  
Our thermal evolution model shows that MH-III may be present today in Titan's  core, confined to a thin ($\approx 10$\,km) outer shell. 
We find that the heat capacity of MH-III is higher than measured values for pure water-ice, larger than heat capacity often adopted for ice-rock mixtures with implications for internal heating.

\section{INTRODUCTION}

Extraterrestrial habitability has always fascinated humanity. The discovery of exoplanets makes the question of habitability of much practical importance.  
Current and future space missions, such as the Transiting Exoplanet Survey Satellite (TESS) and James Webb Space Telescope (JWST), will provide us with spectroscopic atmospheric characterizations. Improving constraints on habitability will provide us with better target filters for future observations. 

Metabolism requires energy, which life obtains in the process of electron transfer by redox chemical reactions \citep{McKay2014,Jelen2016}. One such reaction, analogous to photosynthesis on Earth, and invoked as a possible source of energy for life on Titan, is the production of organics from CH$_4$, followed by release of H$_2$. In addition, for the cryogenic temperatures on the surface of Titan, CH$_4$ replaces water as the surface liquid body, which is essential for cycling nutrients \citep{McKay2016}.
Titan-like worlds may be favorable for habitability being out of harms reach around highly active M-dwarfs \citep{Lunine2009}, and should be common throughout our cosmos considering the ubiquity of water. 
Generalizing beyond Titan-like worlds, CH$_4$, in conjunction with atmospheric O$_2$ and O$_3$, is speculated to be a biosignature \citep{Kaltenegger2010}.
Therefore, modeling the transport of CH$_4$ across the interiors of planets and moons, and its availability at the surface and atmosphere, are of paramount importance. 

Clearly, the data we have for Titan is far superior to what we will have in the foreseeable future for Titan-like exoplanets and exomoons. Therefore, although our aim is more general, this work will use Titan as our primary model object.   
During the epoch of its accretion Titan maintained an inner core of undifferentiated ice-rock mixture \citep[][]{Lunine198761,Barr2010858,Monteux2014377}. The ultimate source of CH$_4$ in present day Titan's atmosphere is this inner core \citep{Lunine198761}. However, this implies that CH$_4$ was able to traverse across the entire depth of Titan to reach the atmosphere.

The moment of inertia estimated for Titan from Cassini gravity measurements suggests that Titan's interior is partially undifferentiated.
The nature of the undifferentiated layer is still unknown. It may be either an ice-rock layer between a rocky core and a water rich outer mantle, or an outer rocky core composed of hydrous silicates \citep{Fortes2012c,lunine10}. If a mixed ice-rock layer indeed exists in the interior of Titan, then a newly discovered high pressure solid solution in the H$_2$O-CH$_4$ binary system called MH-III (also CH$_4$ filled-ice Ih) may hinder the outer transport of internal CH$_4$ into the atmosphere.

The formation of MH-III was first reported in \cite{lovedaynat01}. Contrary to what was previously assumed for the H$_2$O-CH$_4$ system, upon increasing the pressure above about $2$\,GPa, at room temperature, the classical structure H cage clathrate transforms into MH-III, rather then experience phase separation into water ice VII and solid CH$_4$. In this way the solubility of H$_2$O and CH$_4$ is kept throughout the entire pressure range inside Titan. This is also likely true for the much higher pressures spanning water ice mantles of water-rich exoplanets, given that MH-III was found to be stable up to about $100$\,GPa \citep{hirai06}, and temperatures higher than $1000$\,K \citep{Machida2006}. We refer the reader to \cite{loveday01} for a detailed description of the crystallographic nature of MH-III. 

If CH$_4$ is indeed locked in grains of MH-III, scattered within a deep ice-rock layer above the rocky core of Titan, then how much of it may be transported outward depends on the buoyancy of these grains with respect to melt pockets and the solubility of CH$_4$ under such conditions. Such an analysis requires thermophysical data for MH-III at the appropriate conditions.      
In the context of Titan-like exoplanets, previous work on the transport of CH$_4$ across the interior pinpointed the necessary thermophysical data and the lack thereof \citep{Levi2013,Levi2014}. Quantifying the needed thermophysical data is the prime object of this work.

In section $2$ we probe the pressure regime in a possible undifferentiated ice-rock layer inside Titan, to better pinpoint our molecular simulations and the possible role of MH-III. In section $3$ we explain our computational methods. In section $4$ we derive the thermal equation of state and thermal expansivity. Section $5$ is dedicated to the heat capacity. In section $6$ we estimate the thickness of a MH-III enriched layer in Titan's interior, and its carbon content capacity. In section $7$ we use a 1-D thermal evolution model to asses the stability of such a layer. Section $8$ is a discussion, and section $9$ is a summary.

\section{THE PRESSURE REGIME IN THE UNDIFFERENTIATED ICE-ROCK LAYER}

\begin{figure}[ht]
\centering
\includegraphics[trim=0.15cm 12cm 0.01cm 1.0cm , scale=0.60, clip]{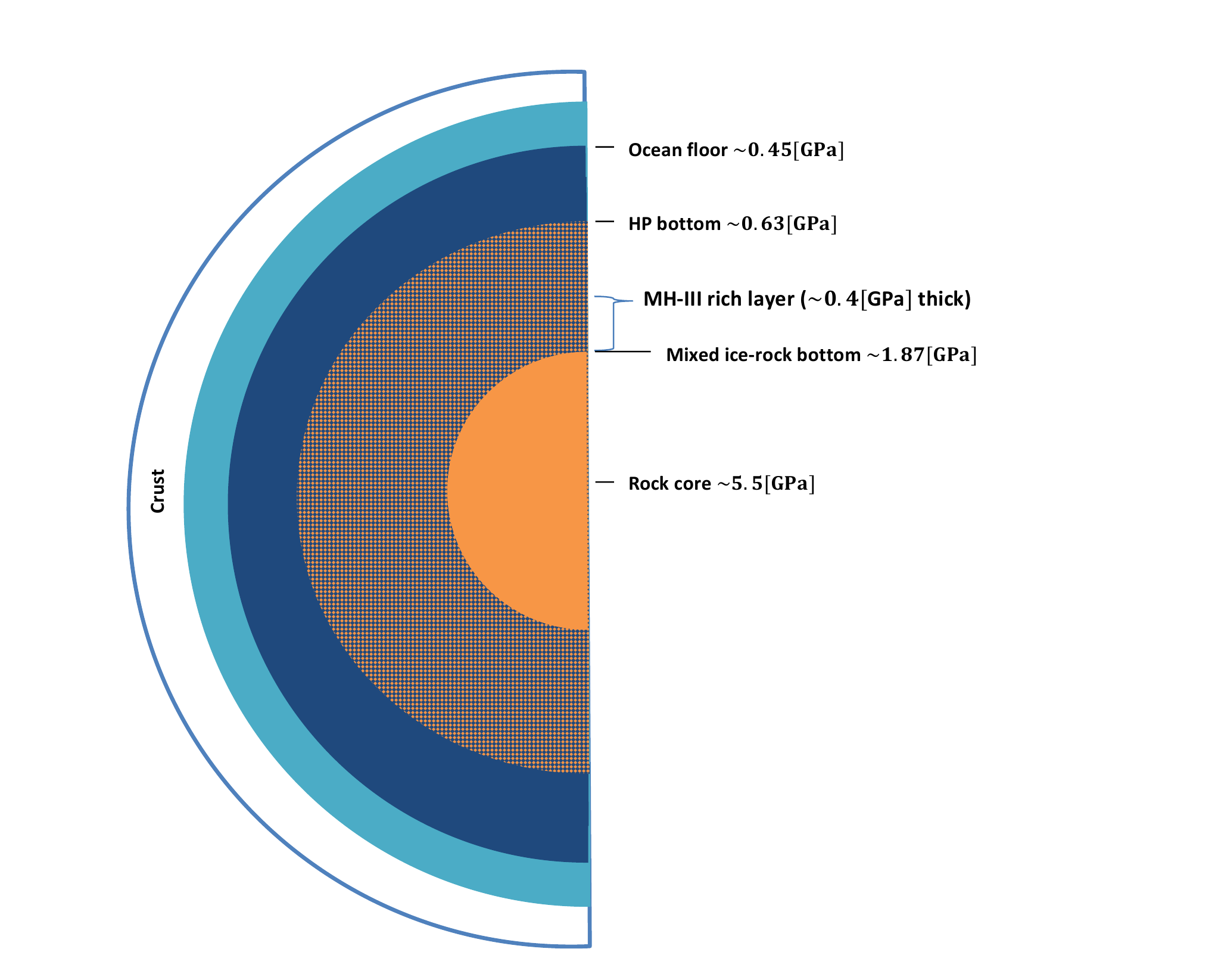}
\caption{\footnotesize{A layered model for Titan: crust, subterranean ocean, high-pressure water ice layer, mixed ice-rock layer, and a rocky core. A layer enriched in MH-III may reside above the core. }}
\label{fig:TitanSection}
\end{figure}

In this section we estimate the pressure at the boundary between a possible internal ice-rock layer and Titan's inner rocky core. We consider three basic constraints for Titan, its radius, mass and normalized moment of inertia (MOI).
We adopt a layered model for the internal structure (see Fig.\ref{fig:TitanSection}). The outermost condensed layer is a shell composed of ice Ih with a mass density of $0.917$\,g\,cm$^{-3}$. We assume for this outer shell a mean thickness of $100$\,km \citep{Nimmo2010}. Underlying the outermost shell is likely a subterranean ocean. \cite{Baland2014} argue for an ocean thickness of less than $100$\,km, whereas \cite{lunine10} list models where the ocean is thicker than $200$\,km. Here, we adopt a width of $200$\,km for the ocean. 

The density range for the subterranean ocean may be inferred from the observational data. It is likely denser than pure water \citep{Iess2012}. \cite{Baland2014} suggest an ocean rich in salts with a mass density in the range $1.275-1.350$\,g\,cm$^{-3}$. \cite{Mitri2014} argue that an ocean denser than $1.2$\,g\,cm$^{-3}$ is inferred from the observed tidal Love number. While the latter is gravitationally compatible with a seafloor composed of ice V, the former range calls for ice VI or VII. \cite{Baland2014} suggest that rock or salt impurities in ice III ($1.16$\,g\,cm$^{-3}$) can increase the overall density and stabilize an overlying dense ocean.
This likely requires strict criteria of size and distribution of rock in the ice, and needs proof. Salty-ice, high-pressure ice with interstitial ions, is another possibility for a density-increasing impurity. However, contrary to ice VII, ions of salt are not very compatible within the smaller voids of ice VI \citep{Frank2006,Journaux2017}, and no information is available for the case of ice V.       
Here we avoid this complication by restricting the ocean density to that of ice V, the high-pressure phase composing the ocean floor, as deduced from our model. 

Beneath the ocean is a high pressure water ice layer, which according to the pressure derived can be composed of either water ice V or VI. We consider the thickness of this layer to be a free parameter, and is here constrained by the observational data.

An undifferentiated ice-rock layer is a possible buffer between the high-pressure water ice layer and a rocky core \citep{Iess2010}.  Its width is taken to be a free parameter, to be estimated by the observational data. If the mass fraction of rock composing this layer is $\chi$, the mixed layer mass density is:
\begin{equation}
\rho_{mix} = \frac{\rho_{hp}\rho_{rock}}{(1-\chi)\rho_{rock}+\chi\rho_{hp}}
\end{equation}
where $\rho_{rock}$ and $\rho_{hp}$ are the rock and high-pressure water ice polymorph densities, respectively.
Below we assume $\chi=0.5$.

The mass density of a rock depends on its composition. Values may range from as little as $2.5$\,g\,cm$^{-3}$ for hydrated rock to as high as $4.5$\,g\,cm$^{-3}$ for anhydrous rock mixed with iron \citep[see discussion in][]{Sohl2003}. \cite{Castillo2010} suggest that carbonaceous and ordinary chondrites are good references for hydrous and anhydrous rock, respectively. A study of carbonaceous chondrites shows they have a wide range of possible grain densities, ranging from $2.42$\,g\,cm$^{-3}$ to $5.66$\,g\,cm$^{-3}$, with an average sample value of $3.44$\,g\,cm$^{-3}$ \citep{Macke2011}. For some specimens bulk density and grain density vary widely due to high porosity. For ordinary chondrites, the average grain density varies from $3.55$\,g\,cm$^{-3}$ to $3.77$\,g\,cm$^{-3}$, for LL and H type, respectively \citep{Consolmagn2006}. High-temperature and pressure in the interior of Titan likely reduce porosity. In any case grain densities should be considered as upper boundary values. \cite{Baland2014} infer a very high density for Titan's core (between $3.39$ and $4.50$\,g\,cm$^{-3}$), implying a high fraction of iron. Therefore, we solve our model for these higher rock mass densities as well, even though they deviate from averaged values.

Finally, the MOI for Titan probably falls in the range of $0.33-0.34$. Non-hydrostatic corrections would favor the lower value.  For an in depth discussion over the MOI we refer the reader to \cite{lunine10}, \cite{Iess2010}, \cite{Baland2014}, and \cite{Mitri2014}. 

We solve for the internal structure of Titan by considering a grid of possible values for the thickness of the high-pressure water ice layer, and the undifferentiated ice-rock layer. This produces contour maps of iso-mass and iso-MOI lines. The solution for the internal structure is obtained by superimposing such two maps, pinpointing the crossing point of the proper contours.

\begin{figure}[ht]
  \begin{minipage}{\textwidth}
  \centering
    \includegraphics[width=.45\textwidth]{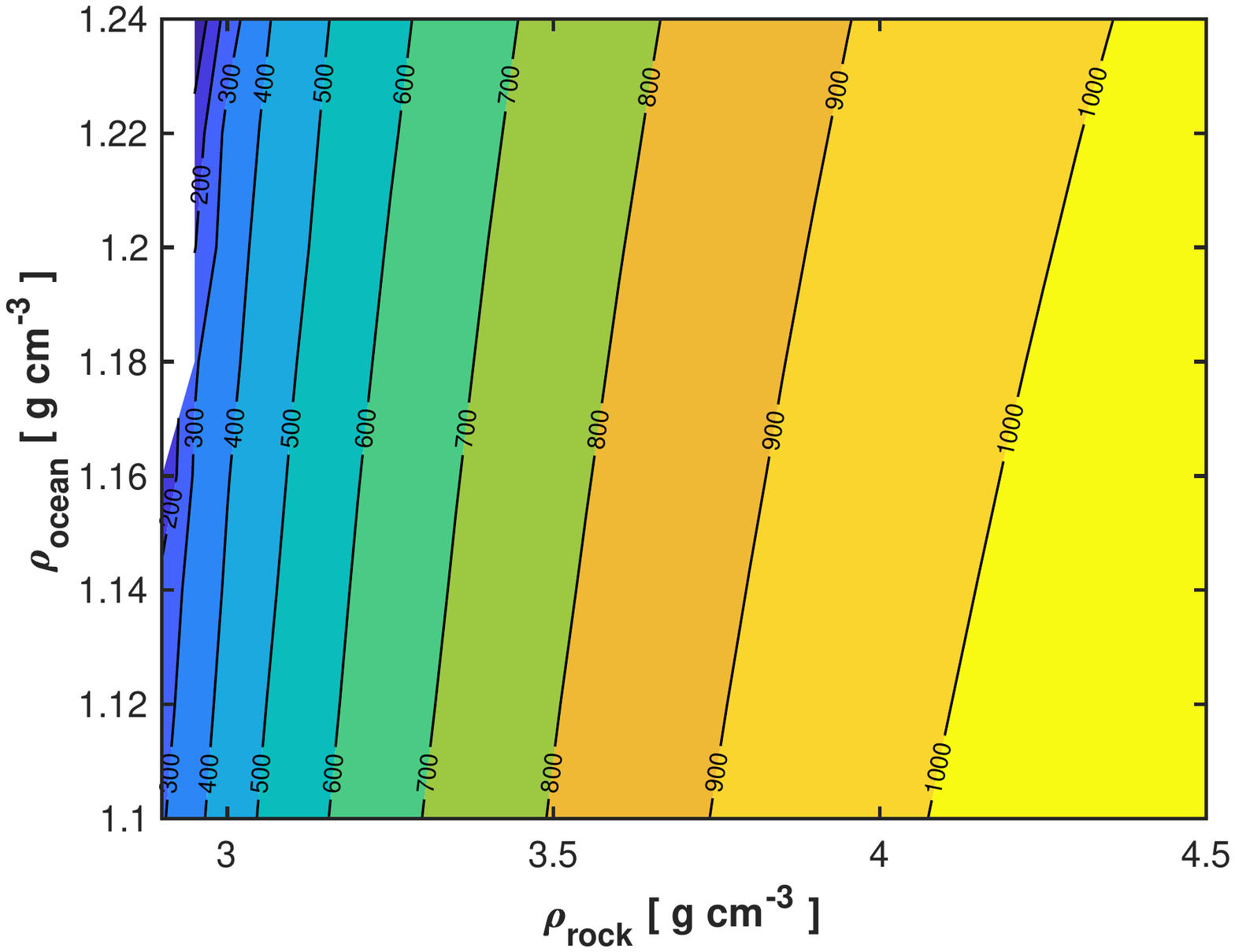}\quad
    \includegraphics[width=.45\textwidth]{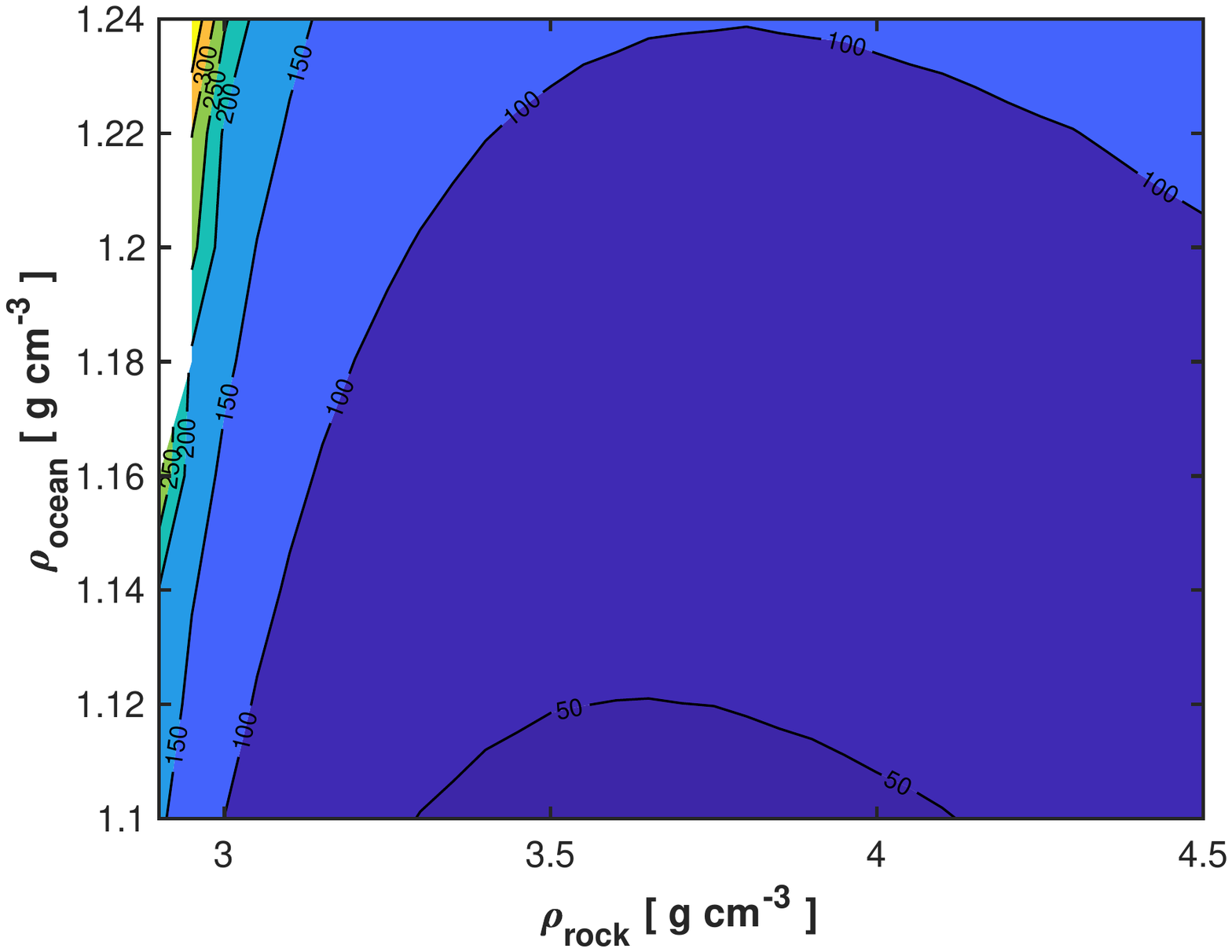}\\
    \includegraphics[width=.45\textwidth]{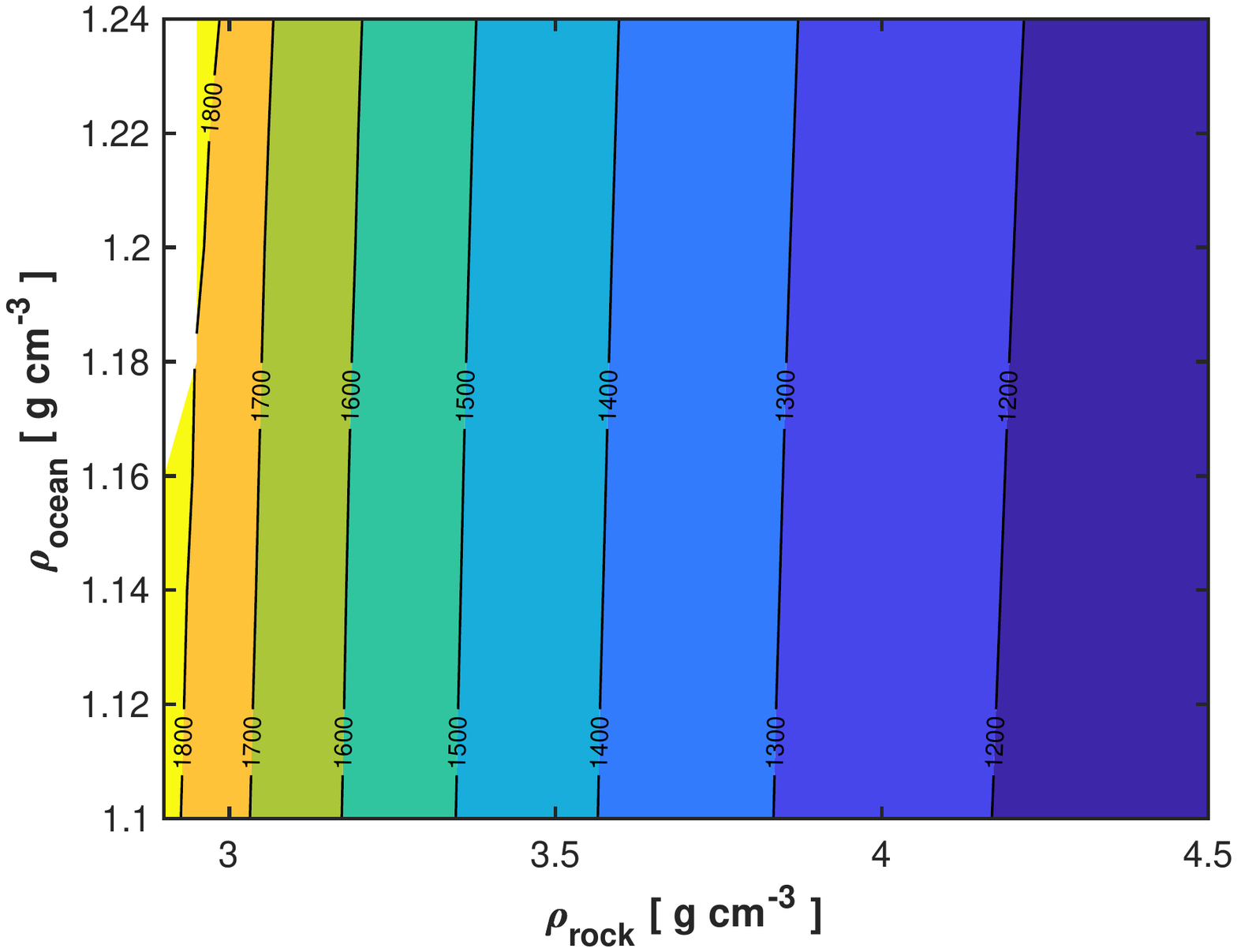}\quad
    \includegraphics[width=.45\textwidth]{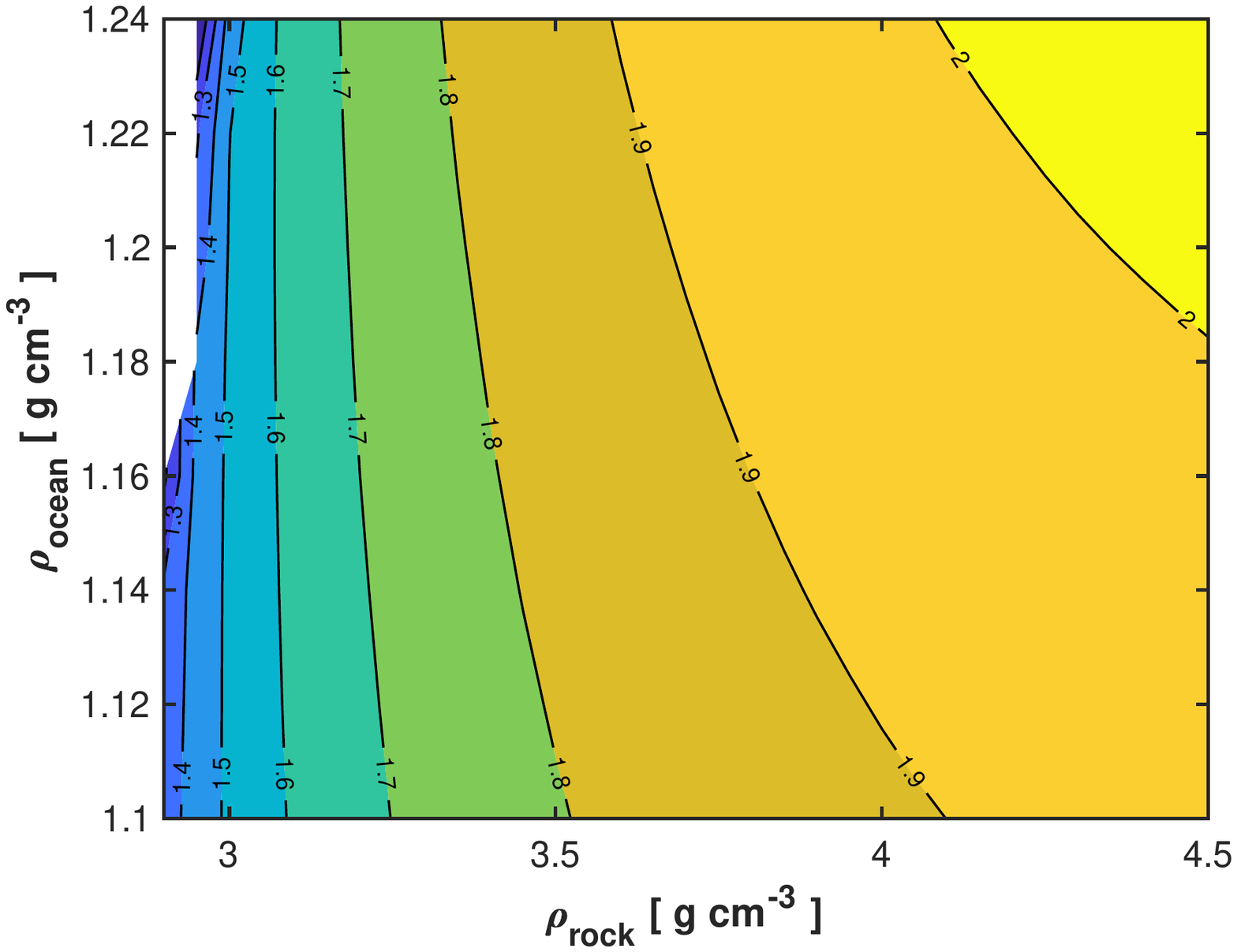}
    \caption{\footnotesize{Thickness [km] of the undifferentiated ice-rock layer \textbf{(upper left panel)}, thickness [km] of the high-pressure ice layer underlying a subterranean ocean \textbf{(upper right panel)}, radius [km] of the rocky core \textbf{(bottom left panel)}, and pressure [GPa] at the boundary between the rocky core and a possible undifferentiated ice-rock layer \textbf{(bottom right panel)}, as a function of rock and ocean mass densities. Data is for a normalized moment of inertia of $0.33$. We assume that the mass fraction of rock in the undifferentiated ice-rock layer is $0.5$. White space is where no solution is found for the given model.}}
    \label{fig:Internal033}
  \end{minipage}\\[1em]
\end{figure}

\begin{figure}[ht]
\centering
\mbox{\subfigure{\includegraphics[width=7cm]{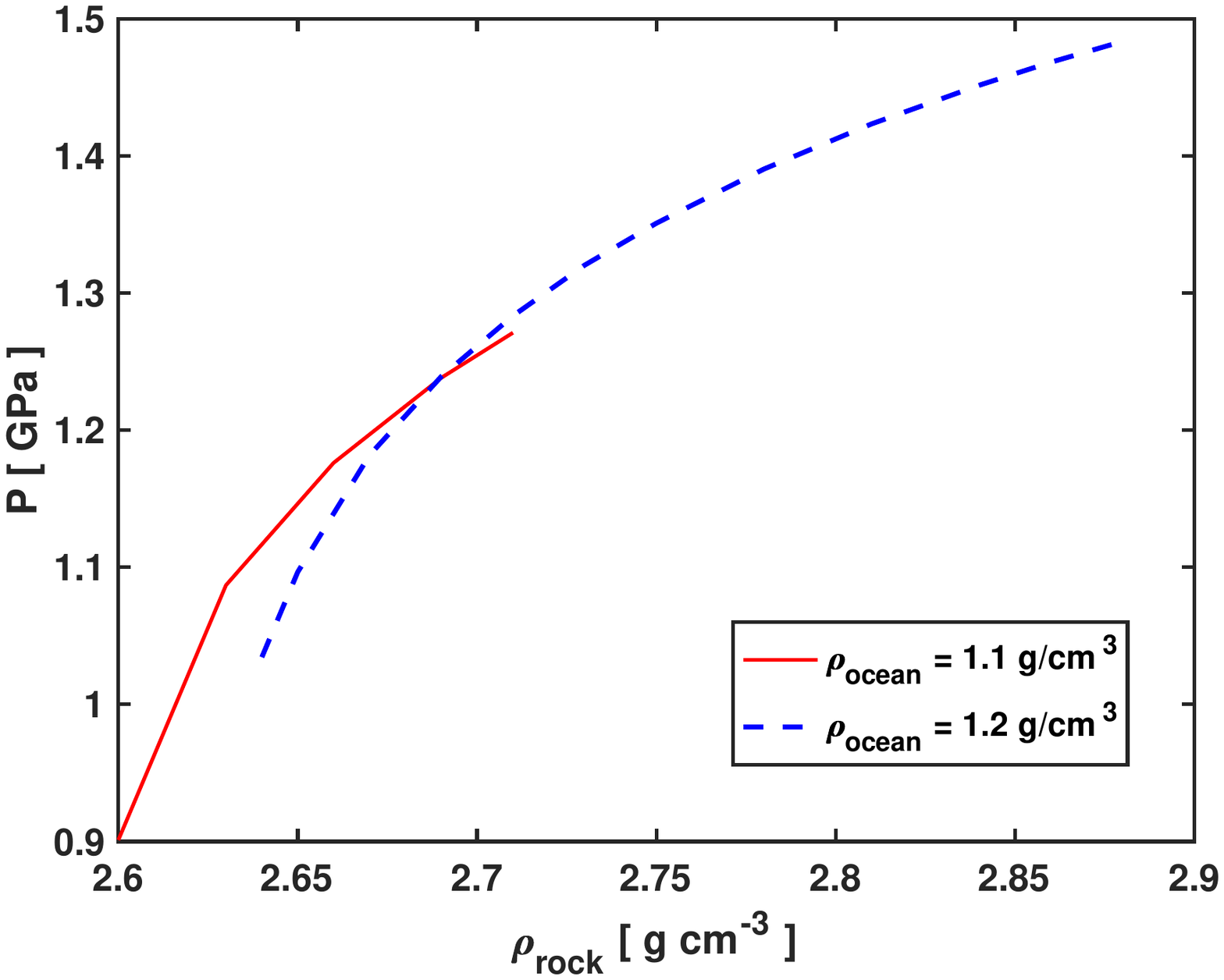}}\quad \subfigure{\includegraphics[width=7cm]{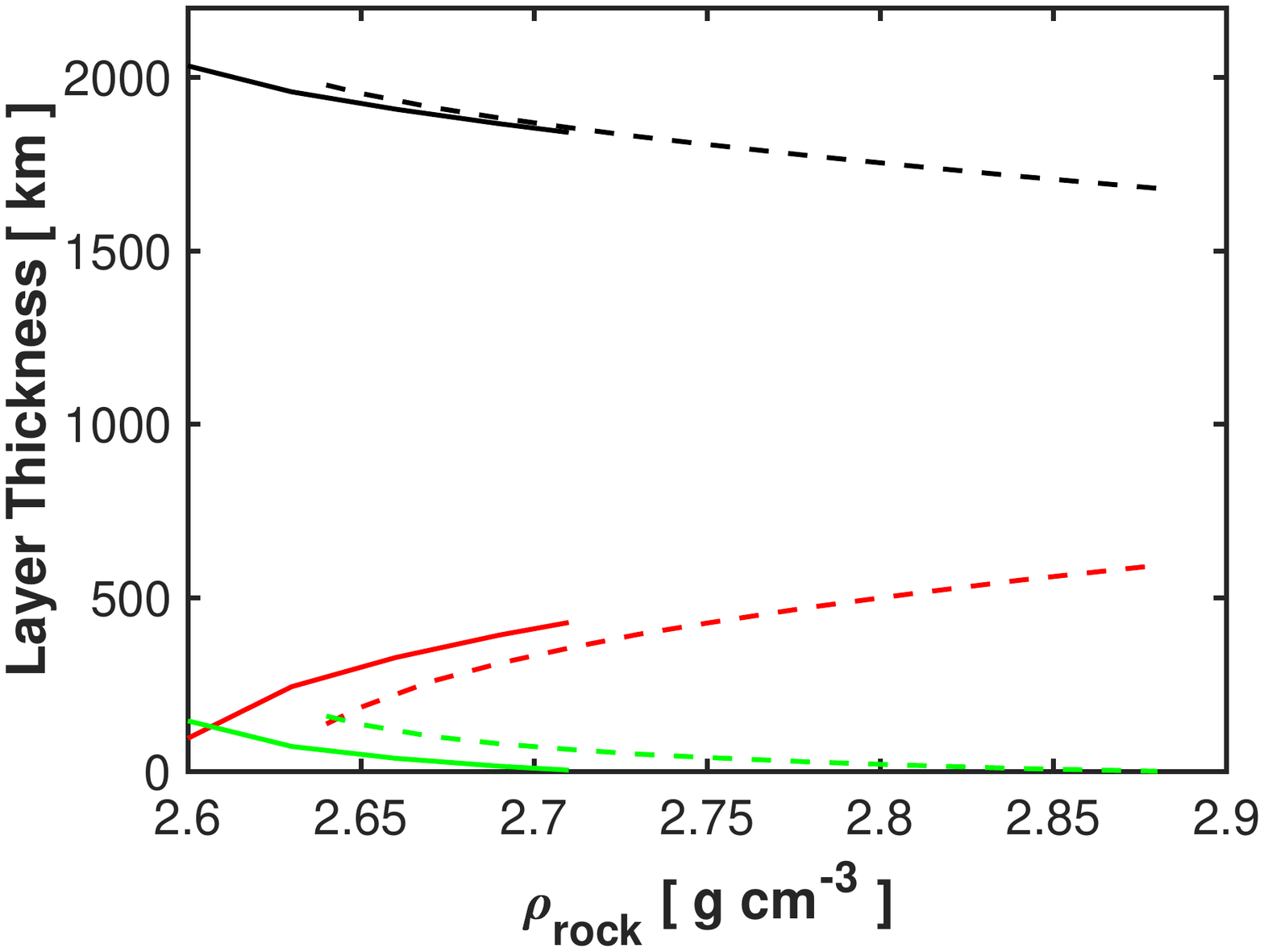}}}
\caption{\footnotesize{\textbf{(left panel)} Pressure at the boundary between the rocky core and a possible undifferentiated ice-rock layer versus rock mass density. \textbf{(right panel)} Titan's interior structure versus rock mass density, where black curves represent the radius of the core ($d_{c}$), red curves represent the thickness of the undifferentiated mixed ice-rock layer ($d_{mix}$), and green curves represent the thickness of the high-pressure ice layer ($d_{hp}$). In both panels the solid curves are for an ocean mass density of $1.1$\,g\,cm$^{-3}$, and dashed curves are for an ocean mass density of $1.2$\,g\,cm$^{-3}$. The normalized moment of inertia (MOI) here is $0.34$.}}
\label{fig:Internal034}
\end{figure}

In Fig.\ref{fig:Internal033} and Fig.\ref{fig:Internal034} we give our solutions for a MOI of $0.33$ and $0.34$, respectively. For a given mass, radius and MOI, increasing the rock mass density requires decreasing the core size and increasing the extent of the mixed ice-rock layer. This results in a monotonic increase in pressure at the bottom of the undifferentiated ice-rock layer. 

The higher MOI commands a body less differentiated. In our model increasing the rock mass density, given a rocky core, is in favor of a lower MOI. This is compensated by thickening of the undifferentiated mixed ice-rock layer and thinning of the high pressure water-rich ice layer which has a lower density. This compensation is more severe the larger is the MOI, therefore resulting in a more constrained solution. 

The possible subterranean ocean represents an outer layer. Therefore, a low mass density for its aqueous solution would favor a low MOI. Hence, the sweeter the subterranean ocean is, the tighter is the  constraint on the rock composition (see Fig.\ref{fig:Internal034}).
     
The result of this interplay in the internal structure yields a distinction between the two moments of inertia. In order to determine whether this distinction is also manifested in the ability to outgas CH$_4$, and the role of MH-III, we need the thermophysical parameters of MH-III. This is the subject of the following sections.

\section{COMPUTATIONAL METHODS}

We studied the equation of state of  MH-III using first-principles electronic structure methods. We performed both static total energy relaxations and molecular dynamics (MD) simulations with the CP2K code.
Simulations were carried out using the CSVR thermostat, which was shown to produce for ice I$_h$ the same vibrational spectrum as that derived with the microcanonical ensemble \citep{Bussi2007}.

Throughout this work we use the quickstep framework within CP2K, with the Gaussian and plane waves mixed bases (GPW). We use the Gaussian basis sets (DZVP-MOLOPT-GTH) from \cite{VandVondele2005}, \cite{VandeVondele2007},  in conjunction with the pseudopotentials (GTH-PBE) of Goedecker, Teter, and Hutter \citep{Goedecker1996,Hartwigsen1998,Krack2005}.
Our system is converged with a cutoff energy of $480$\,Ry.  

We use the exchange functional XC\_GGA\_X\_RPW86 from \cite{Marques2012}, and an LDA local correlation functional of \cite{VWN}, with the non-local van der Waals correlation functional of \cite{Kyuho2010}, adopting a cutoff energy of $160$\,Ry for the latter. 

We use VESTA to produce various super cells ($2\times 1\times1$, $2\times2\times1$ and $2\times2\times2$) from an experimental structure, and AVOGADRO to add the hydrogens since the experiments could not constrain the H positions. We test this procedure by calculating the static energy per supercell, yielding the average energy per atom, with an absolute average deviation of $0.004$\%.  We also derived the static unit cell energy for various \textbf{k}-grids ($\Gamma$, Monkhorst-Pack (MK): 2 2 2, 4 4 4, 6 4 4, 8 4 4, 8 8 8). The difference between $\Gamma$ and MK 8 8 8 was found to be  $0.0002$\,Ha\,atom$^{-1}$.  We also tested for proper energy scaling between the primitive cell and our largest supercell. First, we optimized the unit cell at $0$\,K adopting Macdonald $2$ $2$ $2$ for the \textbf{k}-grid, and derived its associated energy. Then, the optimized crystal was used to build the supercell, whose energy was calculated adopting $\Gamma$ sampling only. Both derived energies are in agreement to $0.0023$\%.

We derived the $0$\,K equation of state (EOS) by optimizing the crystal structure of the unit cell using three different \textbf{k}-point sets: $\Gamma$, Macdonald $2\times2\times2$ and MK $8\times4\times4$. This was used as a double test prior to MD simulations: (1) to check for a normal P-V relation, and (2) check that the optimized cell parameters are in agreement with experimental data given at room temperature.   

Our finite temperature MD was performed on a ($2\times2\times2$) supercell composed of $352$ atoms (each unit cell consists of eight H$_2$O molecules and four CH$_4$ molecules). All simulations carried out on the supercell using $\Gamma$ sampling only. Prior to the MD runs the supercell was optimized at $0$\,K and the resulting cell parameters were used for our finite temperature NVT simulations. In table \ref{tab:CellParameters} we summarize our derived cell parameters, at $0$\,K, for our imposed external pressures.

\begin{deluxetable}{ccccc}
\tablecolumns{5}
\tablewidth{0pt}
\tablenum 1
\tablecaption{Unit cell parameters and volume of MH-III from optimization of the supercell at $0$\,K.  }
\tablehead{
\colhead{$P $} & \colhead{$a $} & \colhead{$b$} & \colhead{$c$} & \colhead{$V$} \\
\colhead{$[GPa]$} & \colhead{$[\AA]$} & \colhead{$[\AA]$} & \colhead{$[\AA]$} & \colhead{$[\AA^3]$} }
\startdata
2 & 4.786 & 8.272 & 8.004 & 316.818 \\ 
\hline
3 & 4.761  & 8.279 & 7.673 & 302.369 \\ 
\hline
4 & 4.674 & 8.186 & 7.591 & 290.374 \\ 
\hline
5 & 4.676 & 8.141 & 7.414 & 282.212 \\ 
\enddata
\label{tab:CellParameters}
\end{deluxetable}

Each MD simulation started with a thermalization run until the pressure changed by less than a few percent. Production runs had $20000$ steps per run, with a time step of $0.5$\,fs. Such a runtime was found adequate for producing frequency spectra from the velocity autocorrelation for pure water systems \citep{French2015}. We further tested for the effect of the time step by performing shorter runs ($10000$ steps) adopting a time step of $0.3$\,fs. The latter runs agree well with the former runs using the larger time step.

Convergence was tested using block sampling. Pressure fluctuations are found to agree with theoretical constraints \citep[see page 341 in][]{lanlif5}. Thermal averages reported in this work used only the last two thirds of each MD run.

\section{THERMAL EQUATION OF STATE}

For the equation of state (EOS) we use the following form,
\begin{equation}
V(P,T) = V(P,0\, K)\exp\left[ \int^T_{0\,K}\alpha(P,T)dT \right]
\end{equation}
where $V$ is the volume per atom, $P$ is pressure, $T$ is the absolute temperature, and $\alpha$ is the volume coefficient of thermal expansion.

For the $0$\,K EOS we use a form of the Murnaghan equation,
\begin{equation}
V(P,0\, K) = V_0\left( \frac{B_0+\tilde{B}_0P}{B_0+\tilde{B}_0P_0} \right)^{-1/\tilde{B}_0}
\end{equation}
where $B_0$ and $\tilde{B}_0$ are the isothermal bulk modulus and its pressure derivative, respectively, at $0$\,K.

For the volume coefficient of thermal expansion we assume a similar form to that in \cite{fei1993},
\begin{equation}
\alpha(P,T) = \alpha_0(T)\left( \frac{B_0+\tilde{B}_0P}{B_0+\tilde{B}_0P_0} \right)^{-\eta}
\end{equation}

\begin{figure}[ht]
\centering
\includegraphics[trim=0.15cm 6cm 0.01cm 7.0cm , scale=0.60, clip]{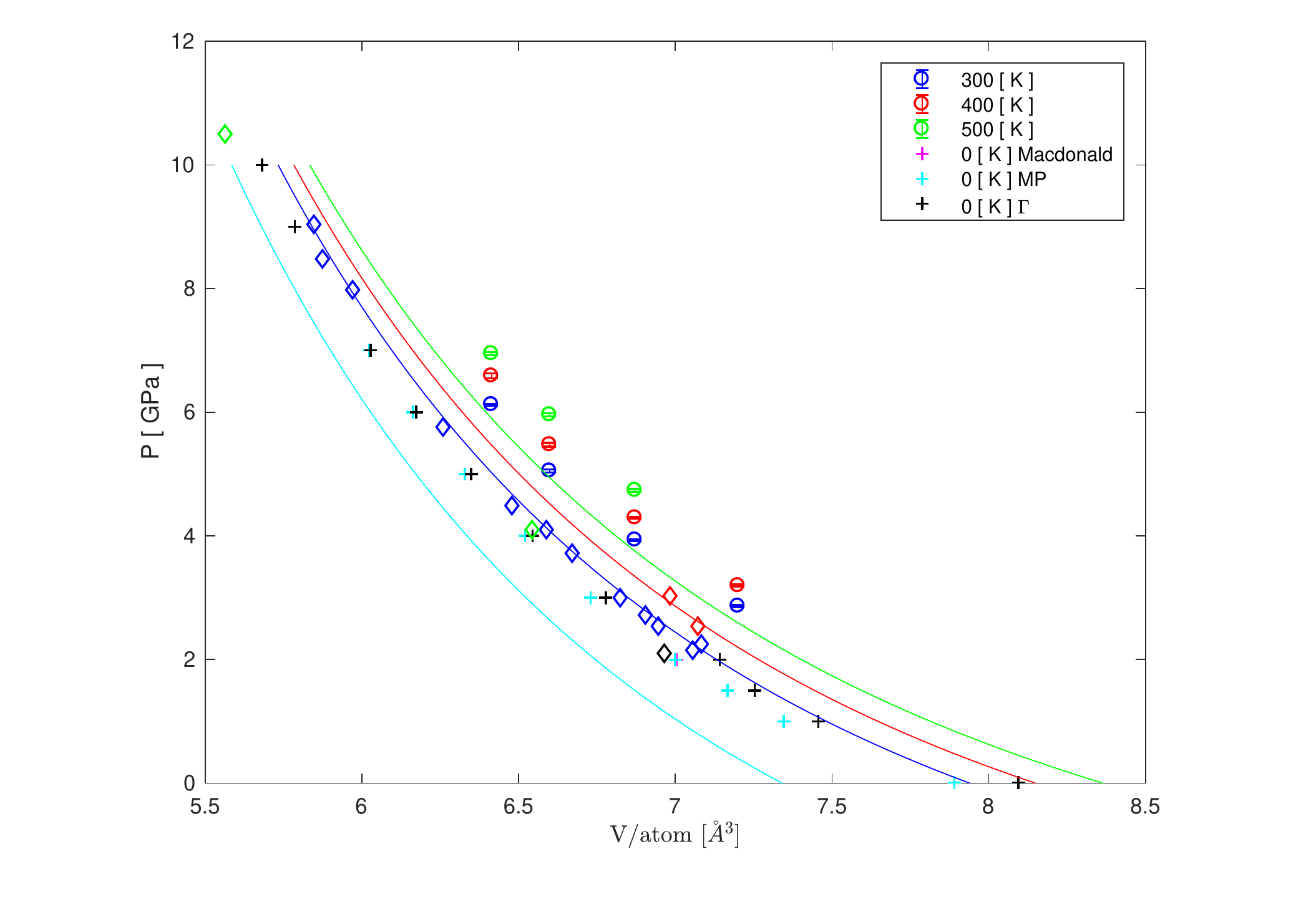}
\caption{\footnotesize{Pressure versus volume per atom for the four isotherms studied in this work. Experimental data: (blue diamonds) room temperature data from \cite{lovedaynat01}, (green diamonds) room temperature data from \cite{hirai2003}, (red diamonds) is data from \cite{Bezacier2014b} at about $400$\,K, (black diamond) is data from \cite{Baumert2005}.  Our simulations at: $500$\,K,  $400$\,K and $300$\,K are the: green, red and blue circles respectively. Black, cyan and magenta crosses are the $0$\,K results of this work using \textbf{k}-grids of: gamma, Monkhorst-Pack $8\times4\times4$ and Macdonald respectively.     }}
\label{fig:EOS}
\end{figure}

In Fig.\ref{fig:EOS} we give our results together with published experimental data.
A comparison with the experimental data shows that our simulations systematically overestimate the pressure by $1$\,GPa. A systematic deviation is likely caused by inaccuracies in the representation of the exchange and correlation functionals, resulting in a systematic error in bond strengths \citep{Lejaeghere2014}. For hydrogen-bonded crystals a systematic shift of approximately $1$\,GPa has previously been reported \citep[e.g.][]{Brand2010}. A systematic error in DFT calculations can be corrected to improve predictability of experimental results \citep{Lejaeghere2014}.        
Indeed, by subtracting $1$\,GPa our results match the experimental data (at $300$\,K and $400$\,K) well. This is used here to benchmark the magnitude of the systematic deviation.
The parameters of the EOS are obtained by a fit to the experimental data and to the data derived in this work after shifting it by $1$\,GPa. Our EOS with the parameters given below fit the data sets with an absolute average deviation of  $0.31$\%.

For $P_0=0$\,GPa we find the following values with $1\sigma$ error bars: $V_0=7.341\pm 0.010$\,\AA$^3/$atom, $B_0=19.72\pm 0.55$\,GPa and $\tilde{B}_0=4.133\pm 0.188$.   

In table \ref{tab:CellParam} we compare between the unit cell parameters at $300$\,K from experiment and our derived values before and after the $1$\,GPa shift in pressure. First the EOS is used in order to determine the unit cell volume at the pressure and temperature of the reference experiment, and then the relation between the cell parameters and the volume, that we have obtained by cell optimization, is used to infer the former. The reported error is a result of the inversion process using the EOS. 

\begin{deluxetable}{cccccccccc}
\tablecolumns{5}
\tablewidth{0pt}
\tablenum 2
\tablecaption{Unit cell parameters of MH-III at $300$\,K. The experimental data (subindex $exp$) at room temperature, is taken from \cite{lovedaynat01} for $3$\,GPa and from \cite{hirai2003} for $4$\,GPa. Original simulation results have subindex $cal$, and shifted data is subindex $sh$.   }
\tablehead{
\colhead{$P $} & \colhead{$a_{exp} $} & \colhead{$b_{exp} $} & \colhead{$c_{exp} $}  & \colhead{$a_{cal} $} & \colhead{$b_{cal} $} & \colhead{$c_{cal} $} & \colhead{$a_{sh} $} & \colhead{$b_{sh} $} & \colhead{$c_{sh} $} \\
\colhead{$[GPa]$} & \colhead{$[\AA]$} & \colhead{$[\AA]$} & \colhead{$[\AA]$}  & \colhead{$[\AA]$} & \colhead{$[\AA]$} & \colhead{$[\AA]$} & \colhead{$[\AA]$} & \colhead{$[\AA]$} & \colhead{$[\AA]$}}
\startdata
3 & 4.746  & 8.064 & 7.845 & 4.809$\pm$0.062 & 8.219$\pm$0.081 & 7.923$\pm$0.152 & 4.746$\pm$0.062 & 8.141$\pm$0.081 & 7.787$\pm$0.152  \\ 
\hline
4 & 4.687 & 7.974 & 7.704 & 4.746$\pm$0.064 & 8.141$\pm$0.083 & 7.787$\pm$0.156 & 4.692$\pm$0.064 & 8.070$\pm$0.083 & 7.678$\pm$0.156 \\ 
\enddata
\label{tab:CellParam}
\end{deluxetable}

\begin{figure}[ht]
\centering
\includegraphics[trim=0.15cm 6cm 0.01cm 7.0cm , scale=0.60, clip]{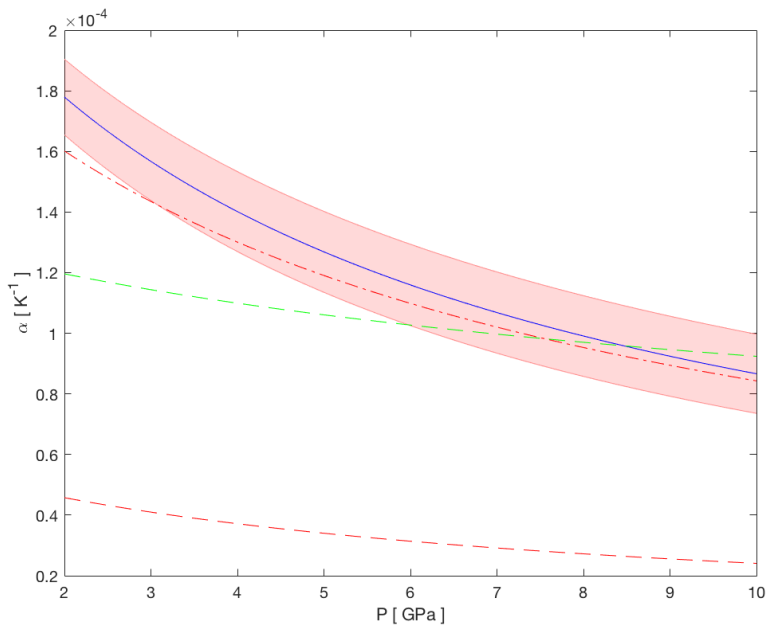}
\caption{\footnotesize{Volume thermal expansion coefficient versus pressure for: MH-III  from this work (solid blue, shaded area is the $1\sigma$ error), water ice VII at $400$\,K from \cite{fei1993}  (dashed-dotted red), water ice VII at $300$\,K from \cite{fei1993}  (dashed red), and water ice VII from \cite{Bezacier2014} (dashed green).}}
\label{fig:ThermalExpansion}
\end{figure}

For the volume thermal expansion coefficient we find that a constant value for  $\alpha_0$ of $2.458\times 10^{-4}\pm (1.68\times 10^{-5})$\,K$^{-1}$, and $\eta = 0.923 \pm 0.121$, best fit the data. An investigation of ice VII in the pressure range of $0-10$\,GPa and temperature range of $300-450$\,K detected no temperature dependence for $\alpha_0$ \citep{Bezacier2014}.  However, an earlier work suggested a linear dependence on the temperature \citep{fei1993}. We note that our modeling approach was shown to produce ice VII lattice parameters that are in good agreement with experimental data \citep[e.g.][]{Futera2018}.  In Fig.\ref{fig:ThermalExpansion} we show our derived thermal expansion coefficient for MH-III and that for ice VII, to aid in comparison between the two phases. For this purpose we fit our suggested equation of state to the data in \cite{Bezacier2014} with an absolute average deviation of $0.25$\%. We adopt the bulk modulus and thermal expansion coefficient of \cite{Bezacier2014}.     

At room temperature the thermal expansion coefficient for MH-III is about $3.7$ times larger than that reported for ice VII in \cite{fei1993}. Although, at $400$\,K this ratio reduces to about $1.06$. Compared with the data reported by \cite{Bezacier2014}, the latter ratio is on average $1.16$. 

At lower pressure H$_2$O-CH$_4$ takes the form of a sI clathrate hydrate.
The pure water ice polymorph in equilibrium with this phase is ice Ih. In the temperature range $100-250$\,K the sI clathrate hydrate to ice Ih volumetric thermal expansion ratio is in the range of $2-4$ \citep{hester2007,feistel06}. Our results, therefore, represent a continuation of this trend to the high pressure MH-III structure as well.

\section{HEAT CAPACITY}

\begin{figure}[ht]
\centering
\includegraphics[trim=0.15cm 6cm 0.01cm 7.0cm , scale=0.60, clip]{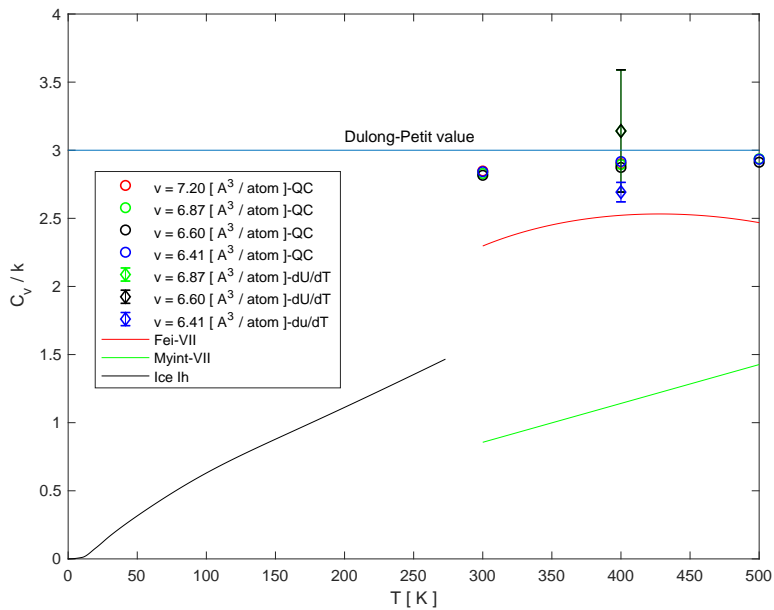}
\caption{\footnotesize{The isochoric heat capacity in terms of degrees of freedom (DOF) per atom versus temperature. Diamond data points and error bars are from the derivative of the internal energy with respect to the temperature (this work). Circle data points are after applying the quantum-mechanical correction using eq.\ref{QuantumCorrCv} (this work). Solid red curve is from data published in \cite{fei1993} for ice VII. Solid green curve is from data published in \cite{Myint2017} for ice VII. Solid black curve is from data published in \cite{feistel06} for ice Ih.}}
\label{fig:CvDOF}
\end{figure}

In Fig.\ref{fig:CvDOF} we plot the isochoric heat capacity per atom, normalized by Boltzmann's constant. This normalization, and the comparison to the Dulong-Petit reference, provides a direct insight into the importance of quantum corrections to the nuclei dynamics in calculating the heat capacity.  Clearly, for pure water ice polymorphs, our temperature range of interest is likely below the Debye temperature. If this is true for MH-III as well, then our classical MD treatment for the nuclei is inadequate for deriving the heat capacity. In this case, it is preferable to treat both electrons and nuclei quantum mechanically, as in \cite{Benoit1998}.  In this work we apply a quantum correction, using the harmonic oscillator approximation, to our classical MD as suggested by \cite{Berens1983}. We derive the velocity autocorrelation function from our simulation, and its Fourier transform, i.e. the vibrational spectrum of the lattice, $g(\nu)$. The latter conforms with the normalization condition,
\begin{equation}
\int_0^\infty d\nu g(\nu) = 1
\end{equation}
The corrected heat capacity, for a system of $N$ atoms and total mass $M$, is then given by,
\begin{equation}\label{QuantumCorrCv}
C_v=3k\frac{N}{M}\int_0^\infty g(\nu)\left( \frac{h\nu}{kT}\right)^2\frac{e^{h\nu/kT}}{\left( e^{h\nu/kT}-1 \right)^2}d\nu
\end{equation}
This approach shows good agreement with experimental data for the case of high-pressure ammonia \citep{Bethkenhagen2013}, in reproducing the equations of state for water ices VII and X \citep{French2015}, and in reproducing the heat capacity of methane \citep{Tingting2012}. The derived values, using this method, are also tabulated in table \ref{tab:aaa}. Our results indicate a low Debye temperature for MH-III. 

In addition, we estimate the isochoric heat capacity by calculating the derivative of the internal energy with respect to temperature using the central finite difference scheme. For a finite central difference scheme the error depends on the third order derivative. We do not have enough data in order to estimate the latter derivative. Therefore, the error in the finite difference scheme is here estimated using the second order derivative, which is the error in the less accurate forward finite difference scheme. Thus, the error reported here is a maximal value. We further consider the error due to statistical errors in the simulated energy using block sampling. However, this error is lower than the former estimated error, except for the case of the atomic volume of $6.41$\AA $^3$\,atom$^{-1}$.

\newpage
\begin{deluxetable}{ccc}
\tablecolumns{3}
\tablewidth{0pt}
\tablenum 3
\tabletypesize{\scriptsize}
\tablecaption{Normalized C$_v$ of MH-III }
\tablehead{
\colhead{$V $} & \colhead{$T $} & \colhead{$C_v/k$}   \\
\colhead{$[\AA^3\,atom^{-1}]$} & \colhead{$[K]$} & \colhead{$$}   }
\startdata
\multirow{2}{4em}{7.20} & 300 & 2.85 \\ 
& 400 & 2.90 \\  
\hline
\multirow{3}{4em}{6.87} & 300 & 2.83 \\ 
& 400 & 2.90 \\ 
& 500 & 2.94 \\
\hline
\multirow{3}{4em}{6.60} & 300 & 2.82 \\ 
& 400 & 2.87 \\ 
& 500 & 2.91 \\ 
\hline
\multirow{3}{4em}{6.41} & 300 & 2.84 \\ 
& 400 & 2.92 \\ 
& 500 & 2.93 \\ 
\enddata
\label{tab:aaa}
\end{deluxetable}

In studying planetary internal structure and formation the isobaric heat capacity is required. 
For MH-III we convert our derived isochoric heat capacity into isobaric using the thermodynamic relation:
\begin{equation}
C_p = C_v+\frac{\alpha^2TV}{\tilde{\kappa}}
\end{equation}
where the volume, $V$, coefficient of volumetric thermal expansion, $\alpha$, and the compressibility, $\tilde{\kappa}$, were estimated in the previous section. 

\begin{figure}[ht]
\centering
\includegraphics[trim=0.15cm 6cm 0.01cm 7.0cm , scale=0.60, clip]{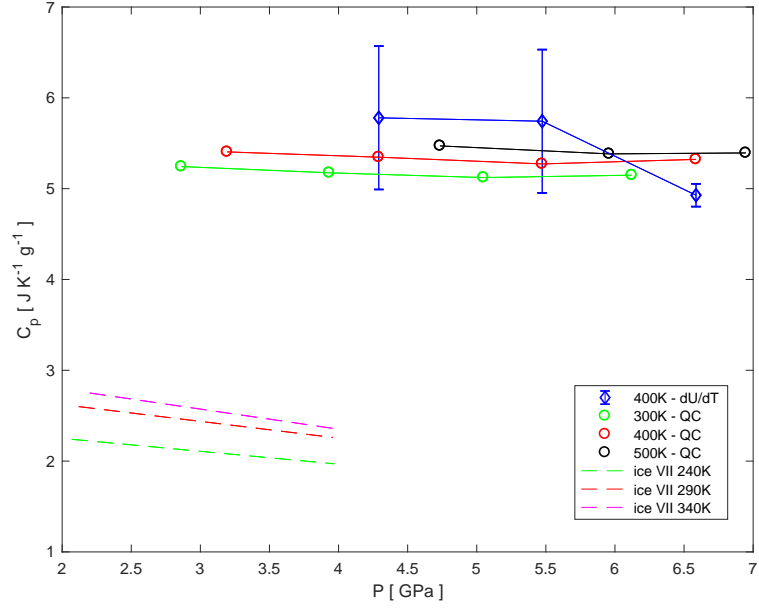}
\caption{\footnotesize{The isobaric heat capacity versus pressure, for various isotherms. Blue diamonds and accompanying errors are from the finite difference derivative of the internal energy (this work). Open circles are data points including the quantum correction (this work). Dashed curves are calculated isobaric heat capacity for ice VII from \cite{Tchijov2004}. }}
\label{fig:CpVP}
\end{figure}

\begin{figure}[ht]
\centering
\includegraphics[trim=0.15cm 6cm 0.01cm 7.0cm , scale=0.60, clip]{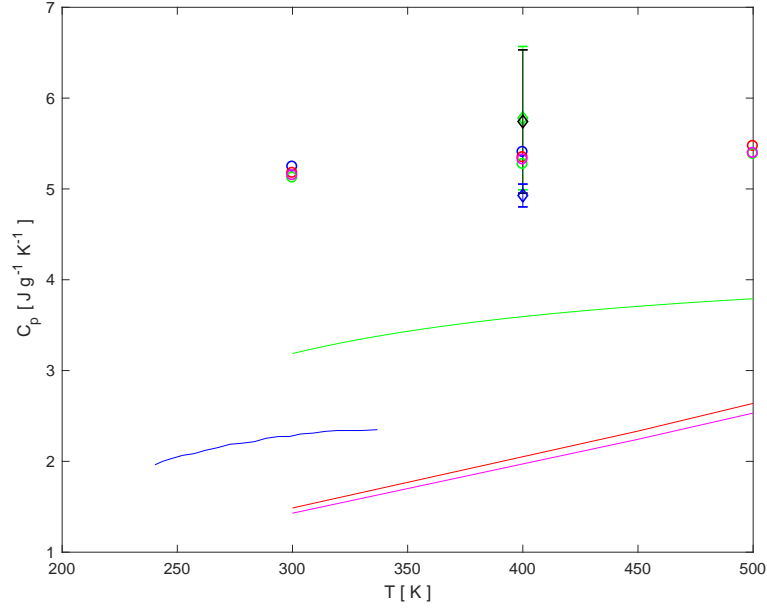}
\caption{\footnotesize{The isobaric heat capacity versus temperature. Diamonds and accompanying errors are from the finite difference derivative of the internal energy from this work: green ($v=6.87$\,$\AA^3$\,atom$^{-1}$), black ($v=6.60$\,$\AA^3$\,atom$^{-1}$) and blue ($v=6.41$\,$\AA^3$\,atom$^{-1}$). Open circles are data points including the quantum correction (this work): blue ($v=7.20$\,$\AA^3$\,atom$^{-1}$), red ($v=6.87$\,$\AA^3$\,atom$^{-1}$), green ($v=6.60$\,$\AA^3$\,atom$^{-1}$) and magenta ($v=6.41$\,$\AA^3$\,atom$^{-1}$). Solid red and magenta curves are for ice VII from \cite{Myint2017}, for $v=6.41$ and $v=6.60$\,$\AA^3$\,atom$^{-1}$, respectively.  Solid blue curve is a calculated isobaric heat capacity for ice VII from \cite{Tchijov2004}. Solid green curve is the isobaric heat capacity for ice VII from \cite{fei1993}}}
\label{fig:CpVT}
\end{figure}

In Fig.\ref{fig:CpVP} we give our derived isobaric heat capacity as a function of pressure, for various isotherms. As for the case of ice VII, the isobaric heat capacity is relatively insensitive to the pressure. In Fig.\ref{fig:CpVT} we give the isobaric heat capacity as a function of the temperature. For clarity, in table \ref{tab:bbb}, we tabulate our derived isobaric heat capacity adopting the quantum harmonic approximation. 

\begin{deluxetable}{cccc}
\tablecolumns{4}
\tablewidth{0pt}
\tablenum 4
\tabletypesize{\scriptsize}
\tablecaption{C$_p$ of MH-III}
\tablehead{
\colhead{$V$} & \colhead{$T$} & \colhead{$P$} &  \colhead{$C_p$}  \\
 \colhead{$ [\AA^3\,atom^{-1}]$} & \colhead{$ [K] $} & \colhead{$[GPa]$} &  \colhead{$[J g^{-1} K^{-1}]$}  }
\startdata
\multirow{2}{4em}{7.20} & 300 & 2.86 & 5.24 \\ 
               & 400 & 3.19  & 5.40\\ 
\hline
\multirow{3}{4em}{6.87} & 300 & 3.93 & 5.17 \\ 
& 400 & 4.29 & 5.35 \\ 
& 500 & 4.73 & 5.47\\ 
\hline
\multirow{3}{4em}{6.60} & 300 & 5.05 & 5.12\\ 
& 400 & 5.47 & 5.27 \\ 
& 500 & 5.96 & 5.38\\ 
\hline
\multirow{3}{4em}{6.41} & 300 & 6.12 & 5.14\\ 
& 400 & 6.59 & 5.32\\ 
& 500 & 6.95 & 5.39\\ 
\enddata
\label{tab:bbb}
\end{deluxetable}

\section{THE PHASE DIAGRAM}

\begin{figure}[ht]
\centering
\includegraphics[trim=1.5cm 6cm 0.01cm 6.5cm , scale=0.65, clip]{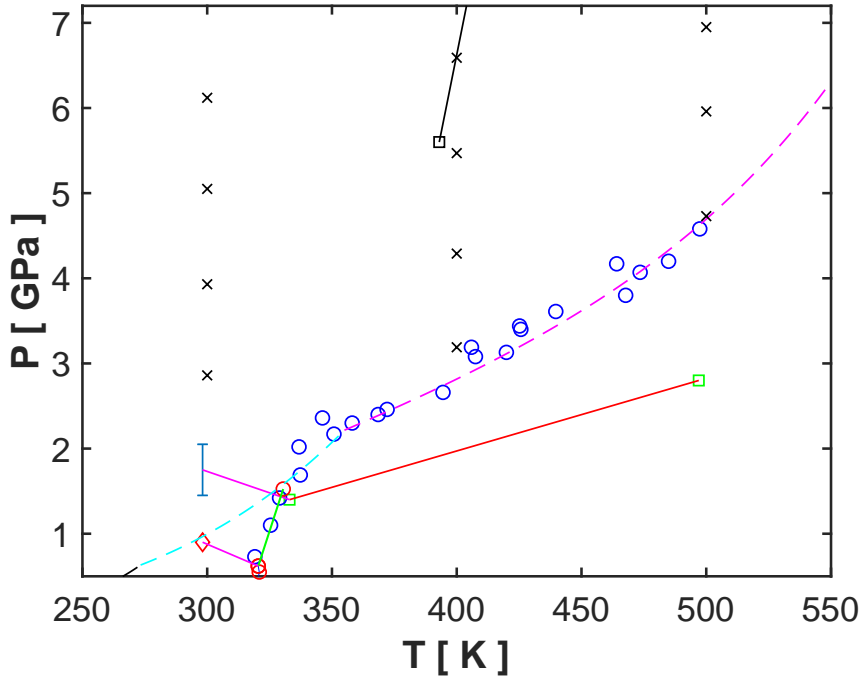}
\caption{\footnotesize{Phase diagram of the H$_2$O-CH$_4$ system. Dashed lines are the melting curves of pure water ice: ice VI (dashed cyan) and ice VII (dashed magenta). Solid green curve is the dissociation curve for sH CH$_4$ clathrate hydrate. Solid red (black) curve is the maximal (minimal) reported dissociation curve for MH-III. Blue circles are data points from \cite{Bezacier2014b} for the dissociation of MH-III and sH CH$_4$ clathrate hydrate.  Solid magenta curves are: the transition from sI to sH (lower curve) and from sH to MH-III (upper curve). Red circles are data points from \cite{dyadin1997}. Green squares are data points from \cite{Kurnosov2006}. Red diamond is a data point from \cite{lovedaynat01}. Error bar encompasses the scatter in the reported data between \cite{Loveday2008}, \cite{lovedaynat01} and \cite{hirai01}. This scatter in the data probably arises from the kinetic inhibition of the phase transition, and from the difficulty in interpreting the neutron diffraction patterns of phases that have disordered protons.   
 Black square is a data point from \cite{Kadobayashi2018}. Black crosses are the sampled points of this work. }}
\label{fig:PhaseDiagram}
\end{figure}

In Fig.\ref{fig:PhaseDiagram} we plot the phase diagram for the H$_2$O-CH$_4$ system, focusing on the likely dissociation curve for MH-III, in the interior of Titan and super-Titan moons.

Three experimental works have been published on the dissociation curve of MH-III with widely varying results \citep{Kurnosov2006,Bezacier2014b,Kadobayashi2018}. \cite{Bezacier2014b} report of a sample at $3.03$\,GPa and $395$\,K where ice VII melted while grains of MH-III remained stable in the liquid water. Therefore, both \cite{Kurnosov2006} and \cite{Bezacier2014b} agree that the thermodynamic stability field of MH-III extends to temperatures higher than the depressed melting temperature of high pressure water ice. However, these two works disagree over the range of this extended stability. \cite{Bezacier2014b} find that the dissociation curve of  MH-III is quite close to the melting curve of pure high pressure water ice. However, \cite{Kurnosov2006} report that the dissociation curve of MH-III penetrates tens of degrees into the pure liquid water regime. The results of \cite{Kurnosov2006} thus imply that CH$_4$ stabilizes filled ice in a way that resembles its stabilizing effect in clathrate hydrates, the latter are known to be stable tens of degrees above the melting curve of water ice Ih. Contrary to previous works \cite{Kadobayashi2018} find that MH-III first separates into ice VII and solid CH$_4$ prior to melting. Another work, though yet unpublished, observed MH-III co-existing with liquid at  $2.32$\,GPa and $370$\,K, about $7$\,K above the melting line of D$_2$O ice VII (Dominic Fortes, Personal Communication).

We note that \cite{Kurnosov2006} worked on the ternary H$_2$O-CH$_4$-NH$_3$ solution. It has been shown, both experimentally and theoretically, that molecules such as NH$_3$ and CH$_3$OH can increase the reactivity of water ice surfaces and act as catalysts to clathrate hydrate formation \citep{Shin2012,Shin2013}. This may be the reason for the more extended stability field for MH-III reported in \cite{Kurnosov2006}. If this is indeed the case, then having $15$\,wt.\% NH$_3$ in their solution suggests that the results of \cite{Kurnosov2006} are less directly applicable to the interior of Titan. This is because NH$_3$ was recently estimated to be only $2-3$\,wt.\% relative to water in Titan's interior \citep{Tobie2012}, in contradiction to earlier and higher estimates \citep{Lunine198761}. The abundance of CH$_3$OH in the interior of Titan is not well constrained.

The temperature immediately above the rocky core is likely kept close to melting conditions \citep{Choblet2017}. Following the melting curve of ice VI, the pressure needed to form MH-III is at its lowest value, of approximately $1.5$\,GPa (see fig.\ref{fig:PhaseDiagram}). Therefore, for a moment of inertia (MOI) of $0.34$, it is not likely that MH-III plays a role in contemporary Titan (see left panel in fig.\ref{fig:Internal034}). However, for an MOI of $0.33$, and for our highest ocean density scenario, a pressure of $1.5$\,GPa is exceeded for $\rho_{rock}>3.05$\,g\,cm$^3$.  
Thus, in this case an undifferentiated ice-rock layer may partially occupy the thermodynamic regime where MH-III is stable. 

For a rock density of $3.5$\,g\,cm$^3$, and ocean density of $1.24$\,g\,cm$^3$, an MH-III enriched sublayer would span a pressure range of approximately $0.35$\,GPa (thickness of $130$\,km) above the rocky core.
If half of the mass of this sublayer is rock, and because the mole ratio of H$_2$O to CH$_4$ in MH-III is $2:1$, then this layer is capable of storing about $7\times 10^{22}$\,mol of CH$_4$, which is about $10^{24}$\,g of CH$_4$.
This is about $3000\times$ the CH$_4$ inventory estimated in Titan's surface and atmosphere \citep{Lorenz2008,Niemann2010}, and represents about $40$\% of the carbon content in Titan's interior \cite[see table $2$ in][]{Tobie2012}.
Earlier work estimated the total mass of CH$_4$ accreted into Titan's core to be $10^{23}$\,g, assuming a hot accretion, and $10^{24}$\,g, assuming a cold accretion \citep{Lunine198761}. Clearly, MH-III is capable of storing a substantial part, if not the lion's share, of the core accreted CH$_4$, possibly hindering its outgassing into the atmosphere. Whether this is the case depends on the stability and thermal evolution of this sublayer which is dealt with in the next section.

\section{THERMAL EVOLUTION AND THE MH-III LAYER}

In this section we quantify the effects a layer enriched in MH-III would have on the thermal evolution of the interior, and consequently on the stability of this layer. We estimate the role of MH-III as a possible CH$_4$ reservoir through time. Again, we will use Titan as an end member test case for our studied worlds. MH-III is a high-pressure ice structure, which if found in the interior, is likely overlying the rocky core. Thus, it is important to first calculate the heat flux out of the core.    

As in \cite{Grasset2000} we assume an initial temperature of $500$\,K for the core, following core overturn. At this stage the high viscosity hinders convective motion and heat is transported conductively \citep{Grasset2000}. The solution for the conductive profile in the rocky core appears in the appendix. This solution is used here to calculate the Rayleigh number for an internally heated core prior to the onset of convection. The appropriate Rayleigh number is \citep{Schubert2001},
\begin{equation}
Ra_{core} = \frac{\alpha_c\rho_cg_cH_cd^5_c}{Cp_c\kappa_c\mu_c}
\end{equation}
where $\kappa_c=7\times 10^{-3}$\,cm$^2$\,s$^{-1}$ is the thermal diffusivity of rock \citep{Yomogida1983}, $Cp_c=9.2\times 10^6$\, erg\,g$^{-1}$\,K$^{-1}$ is the heat capacity of rock \citep{Yomogida1983}, $\rho_c=3.5$\,g\,cm$^{-3}$ is our assumed rock density. This corresponds to a rock thermal conductivity of $2.25\times 10^5$\,erg\,s$^{-1}$\,K$^{-1}$cm$^{-1}$. The rock thermal expansivity coefficient $\alpha_c=2.4\times 10^{-5}$\,K$^{-1}$ is taken from \cite{Kirk1987} based on ultramafic rocks. The dynamic viscosity for the core, $\mu_c$, is here also adopted from \cite{Kirk1987}. For the highest ocean density we have solved for and for our choice for the rock density the core radius is $d_c=1442$\,km. $g_c$ is the acceleration of gravity at mid-core. $H_c$ is the radiogenic heat production rate per unit mass which is adopted from \cite{Grasset2000}. 

Properly scaling the radiogenic heat budget requires the timescale for the core overturn to complete, which as in \cite{Grasset2000} we denote as $t_0$. \cite{Lunine198761} found $t_0\sim1$\,Gyr. As is seen from the phase diagram the presence of MH-III in the undifferentiated rocky core may require higher temperatures in order to achieve melting and core overturn. An upward shift of about $100$\,K in the melting temperature \citep[according to][]{Kurnosov2006}, relative to pure water ice, would increase $t_0$ by $100$\,Myr. We will not consider this effect here due to uncertainties in the initial temperature for the core and the relatively small change to $t_0$.    
Finally, the onset of convection in the rocky core is when the Rayleigh number reaches a critical value here estimated to be $2000$, though this value depends on the boundary conditions \citep[see subsection $7.5$ in][]{Schubert2001}.

If the mole ratio of H$_2$O to CH$_4$ is higher than $2:1$, then any excess water would form ice VI after all CH$_4$ was incorporated into MH-III. The melting temperature of ice VI is likely lower than that for MH-III. Therefore, by being the first phase to melt, and advect excess heat, it is reasonable that the outer boundary temperature for the rocky core is set at the melting temperature of ice VI (here $342$\,K). 

\begin{figure}[ht]
\centering
\mbox{\subfigure{\includegraphics[width=7cm]{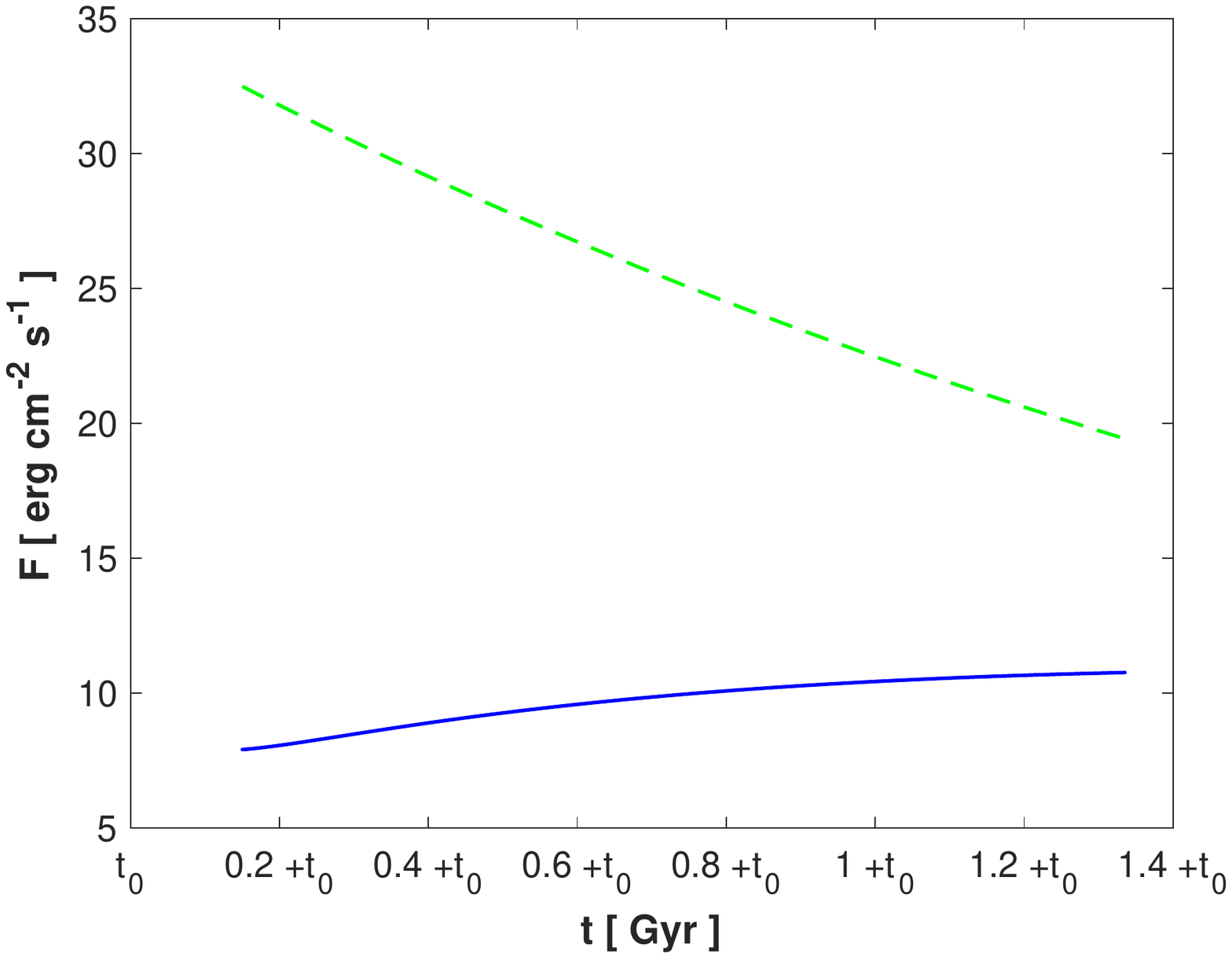}}\quad \subfigure{\includegraphics[width=7cm]{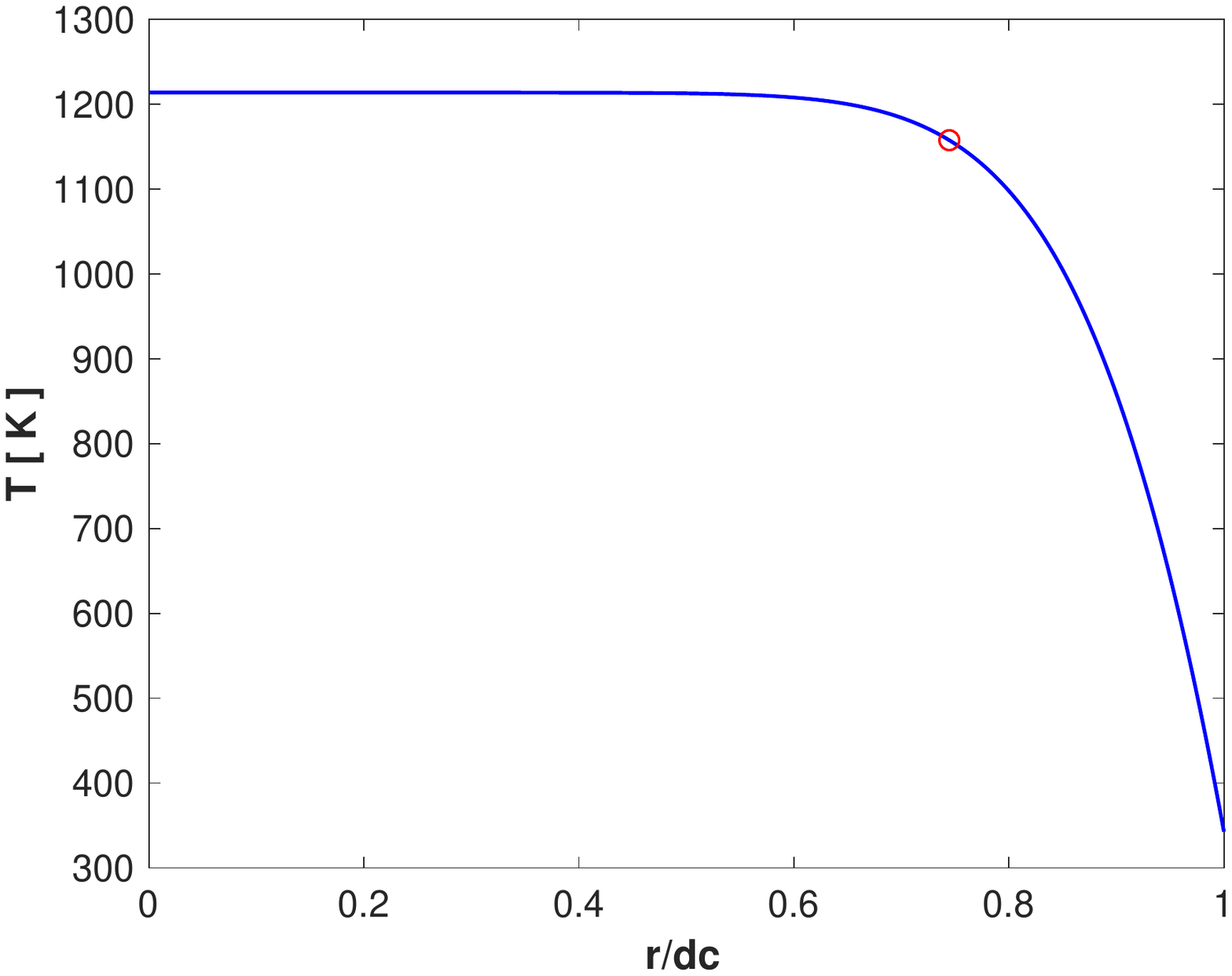}}}
\caption{\footnotesize{\textbf{(left panel)} Heat flux out of the core and into the MH-III enriched layer prior to the onset of convection within the core (solid blue curve). Dashed-dotted (green) is the radiogenic budget scaled to the outer core surface.  \textbf{(right panel)} Conductive thermal profile in the core at the onset of convection. Red circle is the base of the forming stagnant lid.}}
\label{fig:ConductiveProfiles}
\end{figure}

With these assumptions we find that the onset of convection in the core is at $t_{oc}=t_0+1.34$\,Gyr. In Fig.\ref{fig:ConductiveProfiles} we plot the heat flux out of the core, and into a possible MH-III enriched layer, prior to the onset of convection in the core. The scaled radiogenic heat flux is higher than the actual flux indicating a heating stage for the core. We further give in Fig.\ref{fig:ConductiveProfiles} the conductive temperature profile in the core at the onset of convection. The high temperature ($\approx 1200$\,K) will not allow for the survival of MH-III in the inner core at this stage.   

The large temperature gradient in the core, at the onset of convection, corresponds to a very large viscosity contrast ($50$ orders of magnitude). Therefore, a stagnant lid forms, confining convection to the inner part of the rocky core. The temperature difference driving convection in this case is proportional to the rheological temperature \citep{Solomatov1995}. \cite{Davaille1993,Davaille1994} found the following form for this temperature difference,
\begin{equation}
\Delta{T}=-a_{rh}\frac{\mu_c(T_c)}{\left(\frac{d\mu_c}{dT}\right)_{Tc}}
\end{equation}
where $T_c$ is the convection cell temperature, and $a_{rh}=2.24$. \cite{Reese2005} fit their numerical simulations for internally-heated spherical shells to scaling laws with an averaged $a_{rh}=3.2$. 
\cite{Thiriet2019} have shown that 1-D parameterized models for the heat transport can well reproduce data from 3-D thermal evolution models for the stagnant lid regime. They suggest a best fit for $a_{rh}$ of $2.54$. 

By subtracting $\Delta{T}$ from the mid-core temperature we can determine the temperature at the base of the stagnant lid and its thickness, $d_{sl}$, at the onset of convection. We solve for the $a_{rh}$ values suggested above (see table \ref{tab:StagnantLid}). For all cases the initial temperature of the convection cell is $1214$\,K.

\begin{deluxetable}{cc}
\tablecolumns{2}
\tablewidth{2pt}
\tablenum 5
\tablecaption{ Thickness of the stagnant lid at the onset of convection. }
\tablehead{
\colhead{$a_{rh} $} & \colhead{$d_{sl} $(VI)}  \\
\colhead{$ $} & \colhead{$[km]$}  }
\startdata
2.24 & 381   \\ 
\hline
2.54 & 368    \\ 
\hline
3.2 & 344   \\ 
\enddata
\label{tab:StagnantLid}
\end{deluxetable}

In order to derive the heat flux into a possible MH-III enriched layer following the onset of convection in the rocky core we adopt the 1-D model from \cite{Thiriet2019}. The rocky core is divided into three layers, the convection cell, a cold boundary (i.e. the rheological layer), and a stagnant lid. The difference between the heat provided by the convecting core, via the cold boundary layer, and the heat conducted away at the base of the lid yields the stagnant lid evolution with time. The convective heat flow is estimated using the Nusselt number which scales as the Rayleigh number to the power of $\beta$. In this work we adopt the best-fit value of $\beta=0.335$ \citep{Thiriet2019}. Also as in \cite{Thiriet2019} we use a fourth-order Runge-Kutta scheme to advance the convection cell temperature, $T_c$, and the thickness of the stagnant lid with time.   
We solve numerically for the heat conduction within the stagnant lid using an implicit scheme and cartesian coordinates, with $100$ spatial levels.
The initial set up is the stagnant lid thickness and convection cell temperature at the onset of convection, given above. The initial temperature profile along the stagnant lid is derived using the analytical solution given in the appendix.

\begin{figure}[ht]
\centering
\mbox{\subfigure{\includegraphics[width=7cm]{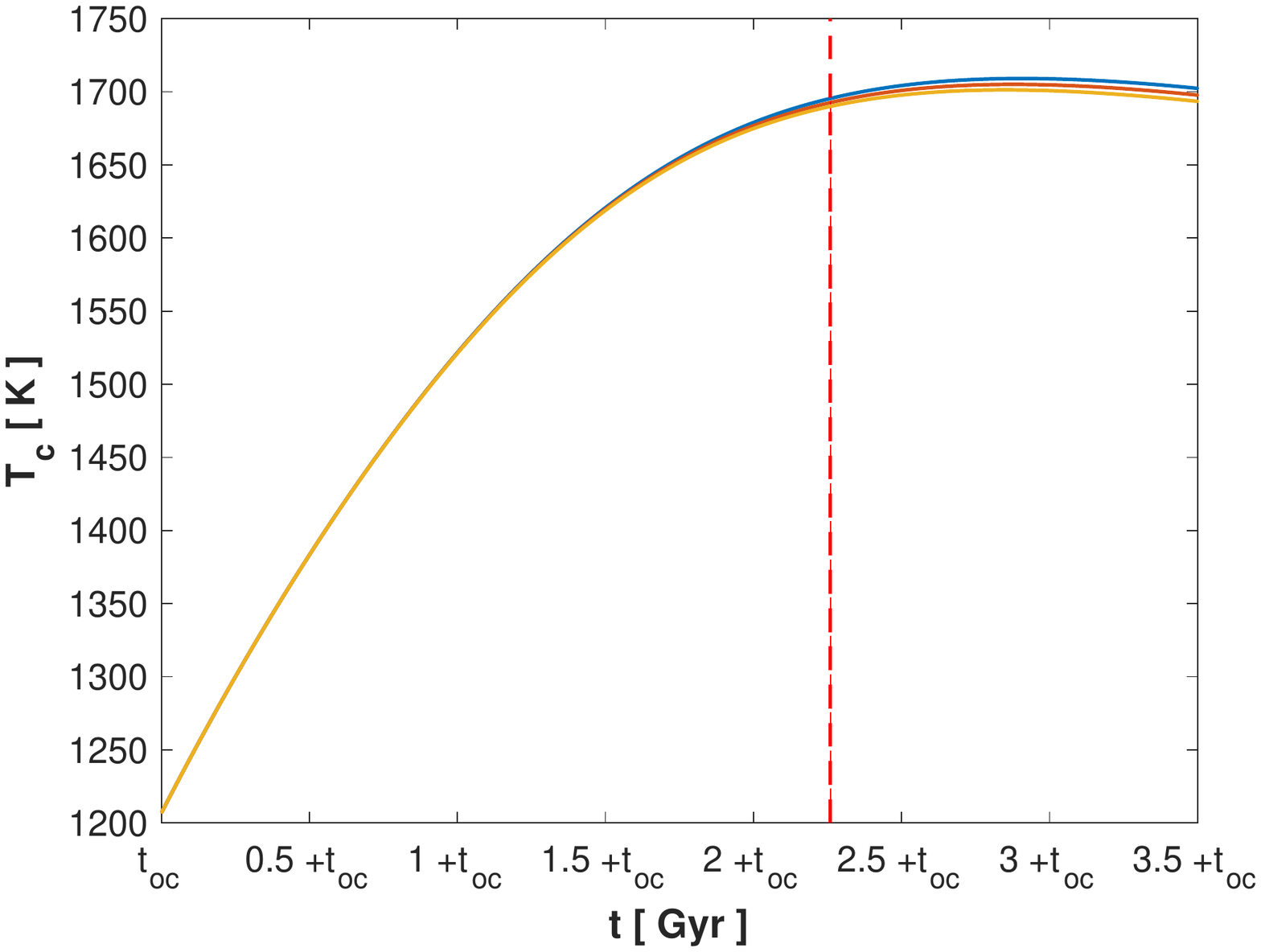}}\quad \subfigure{\includegraphics[width=7cm]{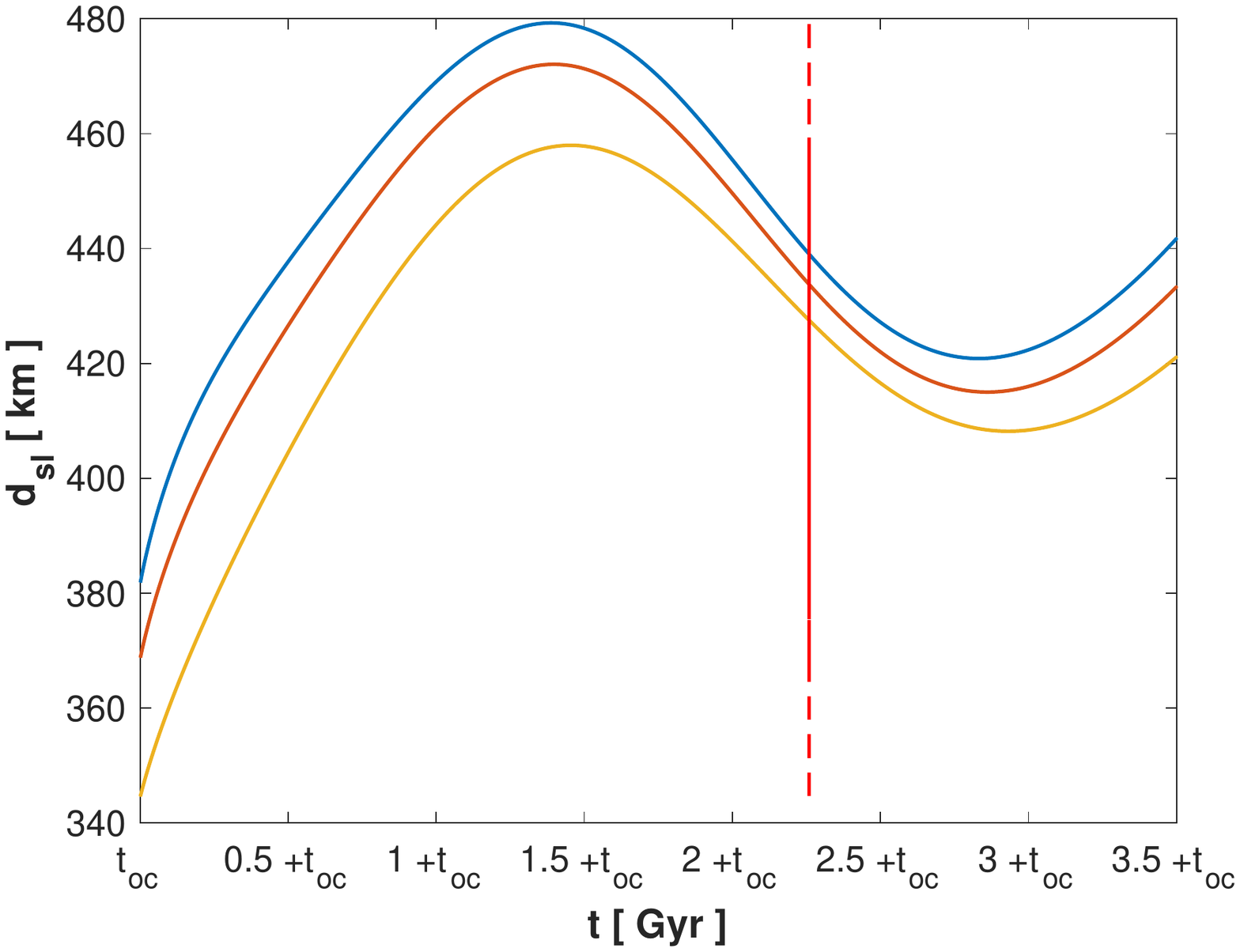}}}
\caption{\footnotesize{\textbf{(left panel)} Temporal evolution of the convection cell temperature in the rocky core, from the onset of convection to the present (the vertical red dashed line marks the present) and projected $\approx 1$\,Gyr into the future. \textbf{(right panel)} Temporal evolution of the rocky core stagnant lid thickness, from the onset of convection to the present (the vertical red dashed line marks the present) and projected $\approx 1$\,Gyr into the future . Solid blue, red and yellow curves are for $a_{rh}$ of $2.24$, $2.54$ and $3.20$, respectively.}}
\label{fig:ConvectionTempLidThickness}
\end{figure}

\begin{figure}[ht]
\centering
\includegraphics[trim=1.5cm 6cm 0.01cm 6.5cm , scale=0.65, clip]{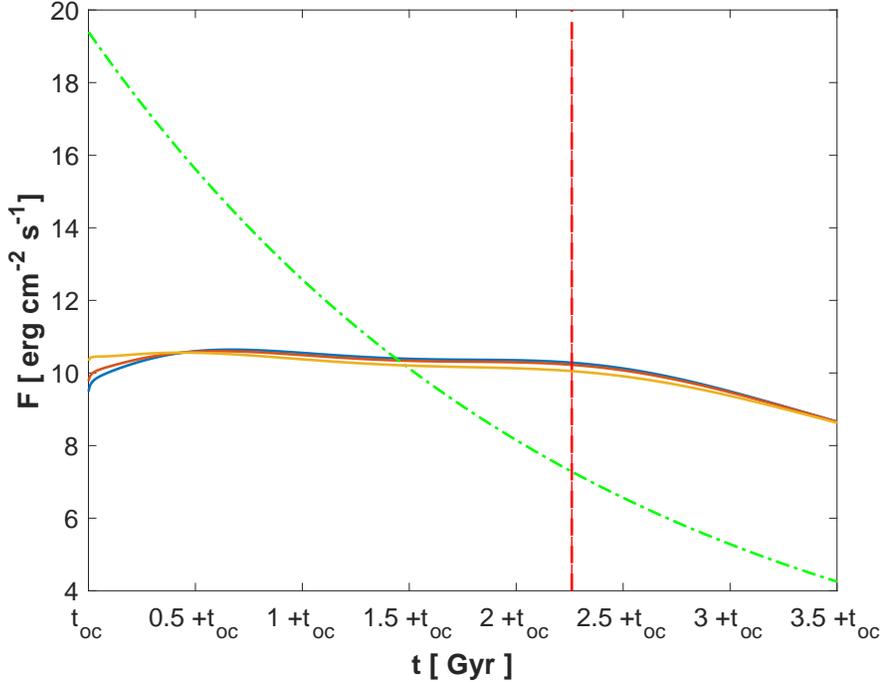}
\caption{\footnotesize{Temporal evolution of the heat flux out of the convective rocky core and into the mixed ice and rock layer,  from the onset of convection to the present (the vertical red dashed line marks the present) and projected $\approx 1$\,Gyr into the future . Dashed-dotted (green) curve is the radiogenic budget scaled to the outer core surface. }}
\label{fig:FluxConvective}
\end{figure}

In Fig.\ref{fig:ConvectionTempLidThickness} we give the evolution with time of the average temperature of the convective part of the rocky core, and of the stagnant lid thickness for the various parameter options from table \ref{tab:StagnantLid}. The projection into the future is a test for a well behaving numerical solution. Fig.\ref{fig:FluxConvective} is the heat flux, following the onset of convection, out of the rocky core and into a possible overlying MH-III enriched layer. Our derived heat flux is not sensitive to the choice for $a_{rh}$.

\begin{figure}[ht]
\centering
\includegraphics[trim=1.5cm 6cm 0.01cm 6.5cm , scale=0.65, clip]{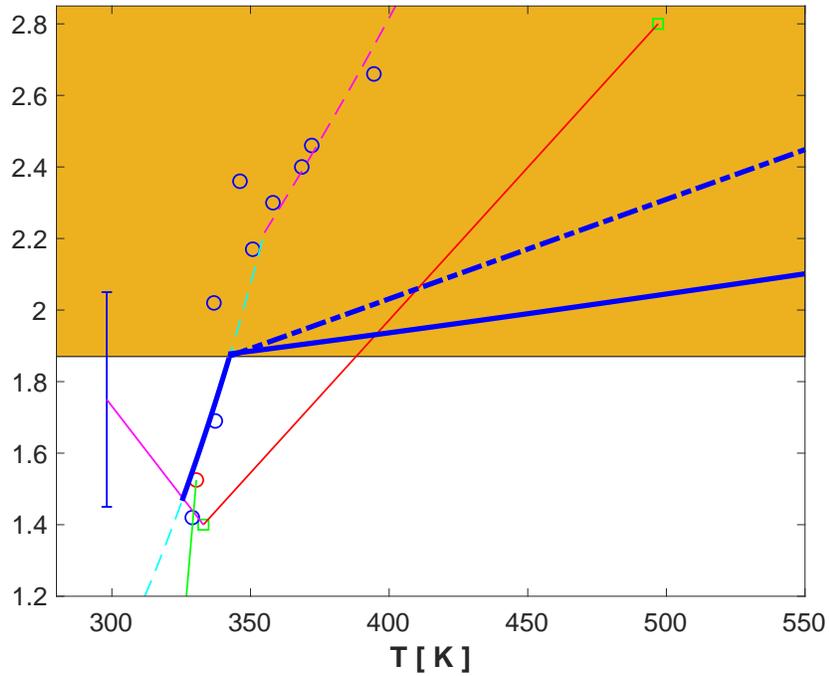}
\caption{\footnotesize{  The thermal profile in the outer part of the core stagnant lid and within the thermodynamic stability field of MH-III in the mixed ice-rock layer. Our derivation for present day Titan (thick solid blue), and for the lower internal temperature in the core from \cite{Grindrod2008} (thick dashed blue). Shaded area represents the rocky core. The reader is referred to Fig.\ref{fig:PhaseDiagram} for a list of all other symbols in the plot. }}
\label{fig:ThermalProfile}
\end{figure}

In Fig.\ref{fig:ThermalProfile} we plot the thermal profile in the stagnant lid and in the mixed ice-rock layer on-top of the H$_2$O-CH$_4$ phase diagram (thick blue curves). Two possibilities emerge for the survival of MH-III, one within the outer part of the rocky core and the other is within the mixed ice-rock layer. 

In addition to the mixed ice-rock layer, and depending on the location of the melting curve of MH-III, it is possible that MH-III survived in a thin layer within the outer part of the rocky core to the present. This though may depend on local concentrations of NH$_3$ and CH$_3$OH. The extent of such a layer depends on the core temperature, and thus the radiogenic budget. For our calculation this layer is $12$\,km wide. For the lower core temperature ($\approx 1000$\,K) from \cite{Grindrod2008} this layer is $39$\,km wide. Assuming the volume fraction of MH-III in this layer varies between $1-10$\%, we find this layer can hold $7\times 10^{19}-7\times 10^{20}$\,mol of CH$_4$. This is equivalent to $3-30\times$ the surface and atmospheric budget of CH$_4$ \citep{Lorenz2008,Niemann2010}. For the lower core temperature of \cite{Grindrod2008} this reservoir is equivalent to $9-90\times$ the surface and atmospheric budget of CH$_4$. If the melting temperature of MH-III is closer to that reported by \cite{Bezacier2014b} then this reservoir is likely negligible, hence leaving MH-III as a possible CH$_4$ reservoir only within the mixed ice-rock layer.       

In order to estimate the mode of heat transport across the mixed ice-rock layer we note that the high-pressure ice cold boundary layer (i.e. high-pressure ice to subterranean ocean interface, see Fig.\ref{fig:TitanSection}) is confined to the melting curve of high-pressure ice \citep{Choblet2017}. The melting curve of ice V has a gradient of approximately $60$\,K\,GPa$^{-1}$. Adopting the thermal conductivity for ice VI from \cite{Chen2011} yields a flux of $0.2$\,erg\,cm$^{-2}$s$^{-1}$. The latter value is far less than our estimated heat flux out of the core. Therefore, melting of ice is a prominent mode of heat transport, as reported by \cite{Choblet2017} and \cite{Kalousova2018}, both during the conductive and convective phase of the core. Thus, it is likely that the thermal profile in the ice-rock layer above the core follows the melting curve of ice VI, the phase with the lowest melting temperature (see thick blue curve in Fig.\ref{fig:ThermalProfile}). Hence, the stability field of MH-III spans about $0.4$\,GPa above the core for our case of study, which as stated in the previous section, suggests this phase may be a major reservoir for CH$_4$.  

\begin{figure}[ht]
\centering
\includegraphics[trim=1.5cm 8cm 0.01cm 6.5cm , scale=0.70, clip]{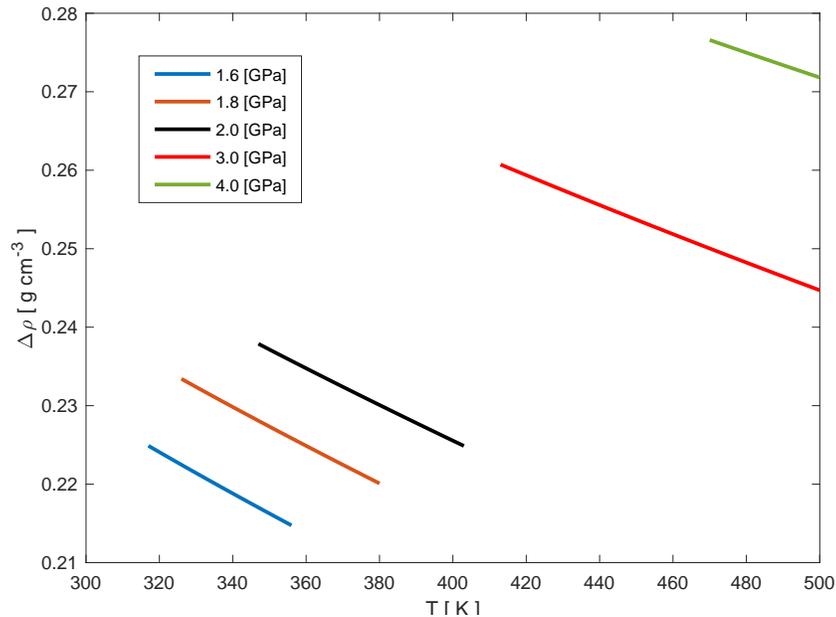}
\caption{\footnotesize{The difference in density between pure liquid water (from \cite{wagner02}) and MH-III (this work), for various isobars spanning part of the thermodynamic stability field of MH-III.}}
\label{fig:densitydiff}
\end{figure}

For melting to be an efficient coolant it ought promote positively buoyant hot plumes. The density of ice VI is indeed higher than that of liquid water, promoting buoyancy and melt extraction. In order to consider the effect of MH-III we plot in Fig.\ref{fig:densitydiff} the density difference between that for pure liquid water (taken from \cite{wagner02}) and for MH-III (derived in this work). 
Liquid water is denser than MH-III for the P-T conditions of interest for modeling Titan and possibly for super-Titans as well. This was observed experimentally at $3.03$\,GPa and $395$\,K, where ice VII melted while the grains of MH-III migrated to the upper part of the vessel where the sample was held \citep{Bezacier2014b}. 

\begin{figure}[ht]
\centering
\mbox{\subfigure{\includegraphics[width=7cm]{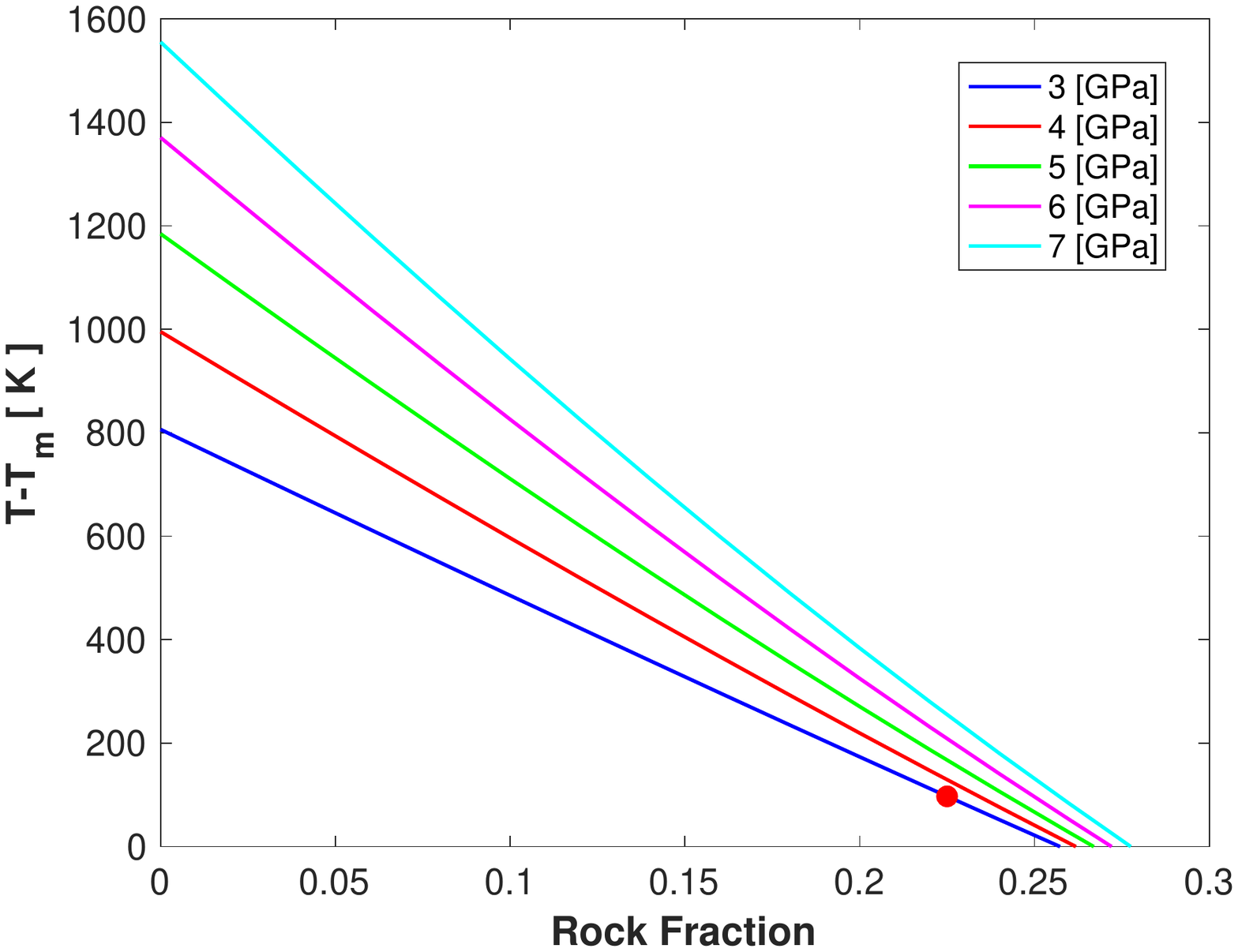}}\quad \subfigure{\includegraphics[width=7cm]{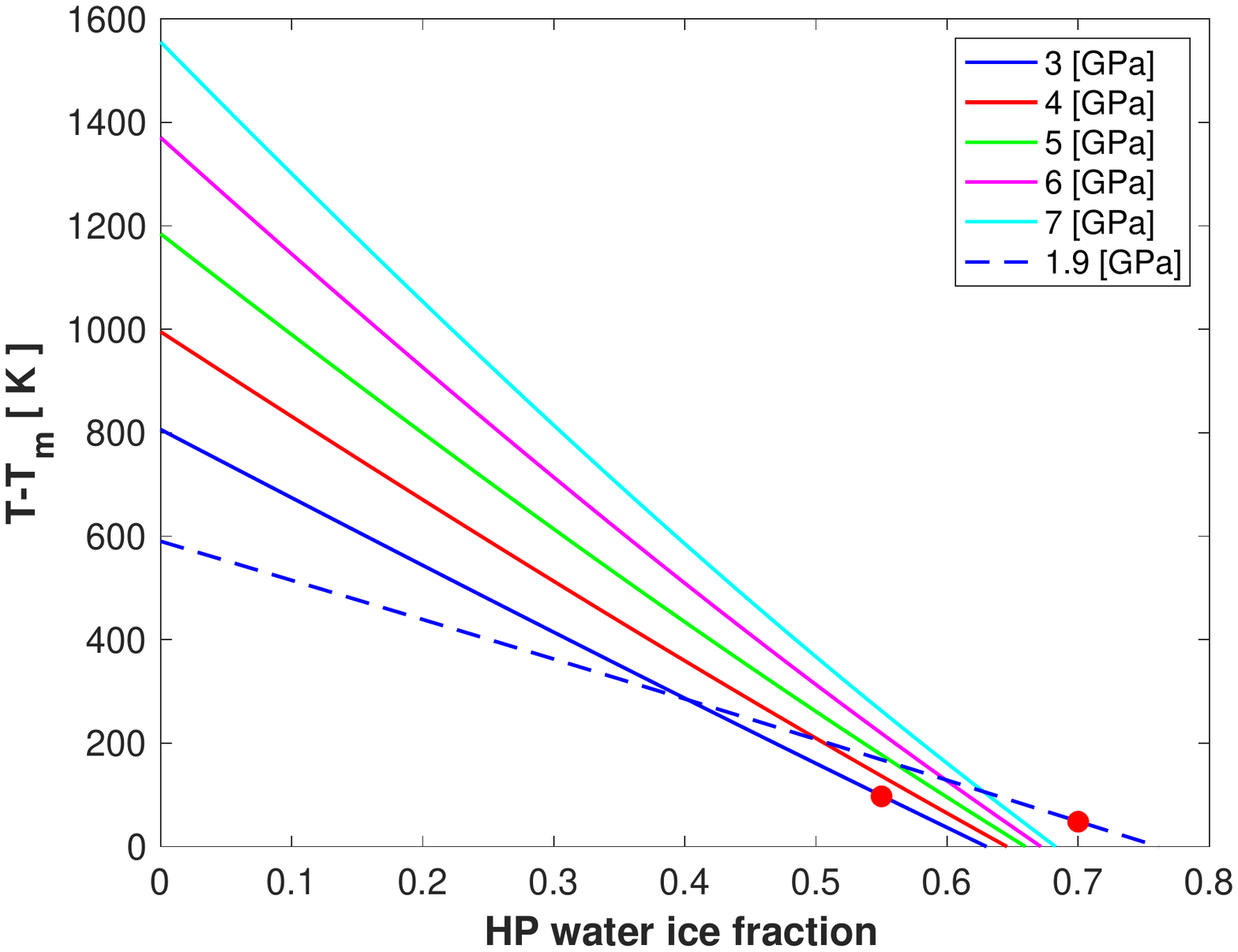}}}
\caption{\footnotesize{\textbf{(left panel)} Excess temperature, above the melting point, needed to make liquid water positively buoyant relative to a mixture of MH-III and rock ($\rho_{rock}=3.5$\,g\,cm$^{-3}$). Solved for various isobars. \textbf{(right panel)} Excess temperature, above the melting point, needed to make liquid water positively buoyant relative to a mixture of MH-III and high-pressure water ice (dashed curve for ice VI, solid curves for ice VII) . Solved for various isobars. Red dots mark the upper boundary reported for the stability field of MH-III by \cite{Kurnosov2006}. }}
\label{fig:ExcessTemp}
\end{figure}

In a water-rich system (if H$_2$O:CH$_4$ mole ratio $>$2:1 ) both ice VI and MH-III would form, or ice VII and MH-III, depending on the pressure. Because ice VI likely has a lower melting temperature than MH-III, then melting on the outer surface of the rocky core would tend to aggregate solid MH-III on top of the melt, consequently hindering the formation of hot positively buoyant plumes. In this case the melt would further increase in temperature until it becomes positively buoyant. In Fig.\ref{fig:ExcessTemp} we plot the excess temperature (temperature above the melting temperature of pure high-pressure water ice) required in order for liquid water to become positively buoyant relative to a MH-III and rock, and MH-III and pure high-pressure ice (VI or VII), composition. The density of ice VI and VII near their melting curve is taken from \cite{Bezacier2014}  and \cite{Frank2004}.  

If the mass fraction of rock ($\rho_{rock}=3.5$\,g\,cm$^{-3}$) is higher than $\approx 0.3$, then the melt is always positively buoyant. If rock is replaced with ice VII (more likely for super-Titans), then this threshold increases to $\approx 0.65$, and to $\approx 0.75$ for the case of ice VI. Therefore, the mode of transport of CH$_4$ across an MH-III enriched layer becomes dependent on the composition of the layer. If the composition is such that the melt is always positively buoyant, then the upwelling melt would carry the fraction of CH$_4$ which can dissolve within the melt. However, if the composition of the layer is such that the melt experiences excess heating in order to become buoyant, then during its upwelling it may dissociate MH-III that it comes in contact with, thus carrying not only the dissolved fraction of CH$_4$, but also CH$_4$ as a separate phase. For example, if the rock mass fraction is higher than $0.26$, or possibly $0.22$, (see $3$\,GPa isobar) then MH-III in the path of a hot plume may be stable, and CH$_4$ migrates as a dissolved component. However, if the mass fraction of rock is less than $0.22$, than the excess temperature of the melt would also dissociate MH-III along its path. The latter value would become closer to $0.26$ for the dissociation conditions of MH-III reported by \cite{Bezacier2014b}. For the case of super-Titans, if the mass fraction of ice VII is less than $\sim 0.5$, hot plumes at the core boundary would dissociate MH-III causing rapid outward migration of CH$_4$, perhaps into the atmosphere.

\section{DISCUSSION}

MH-III is an important phase within the H$_2$O-CH$_4$ binary system that should be considered when modeling water-rich bodies. Its thermodynamic stability field is very wide, overlapping those of ice VI, VII and X. In addition it has a high capacity for storing CH$_4$ (H$_2$O:CH$_4$ mole ratio of $2$:$1$). We show that MH-III may exist in the interior of Titan for part of the parameter space describing Titan's inferred internal structure. However, this phase is likely dominant in the interior of super-Titans owing to the higher pressures reached within their water-rich ice mantles. This has two interesting consequences, (1) it may further constrain Titan's interior models, (2) it may create a dichotomy breaking analogies between Titan and super-Titans. 

We describe two modes for the outward transport of CH$_4$ across a MH-III enriched layer, either as a dissolved component within a buoyant melt, or largely as a separate phase (in addition to partially dissolved) if the melt is of a high enough temperature to dissociate MH-III in its path. These two modes represent different CH$_4$ transport efficiencies if the solubility value is low. The solubility of CH$_4$ when in equilibrium with MH-III is not known. If it is low, and dissolution of CH$_4$ is the prime mode of transport, then melt extraction may not be an efficient mechanism for the outward transport of CH$_4$ out of an MH-III enriched layer. In this case the abundant presence of CH$_4$ at the surface and atmosphere of Titan makes a high moment of inertia ($0.34$) more likely, since MH-III will not be stable in Titan's interior in this case. Another possibility is that the composition of the ice-rock mixed layer, assumed in our model, is poor in rock ($\lessapprox 0.2$ in mass fraction) resulting in a buoyant melt that is hot enough to dissociate MH-III along its path.         

\begin{figure}[ht]
\centering
\includegraphics[trim=0.15cm 5cm 0.01cm 5.0cm , scale=0.60, clip]{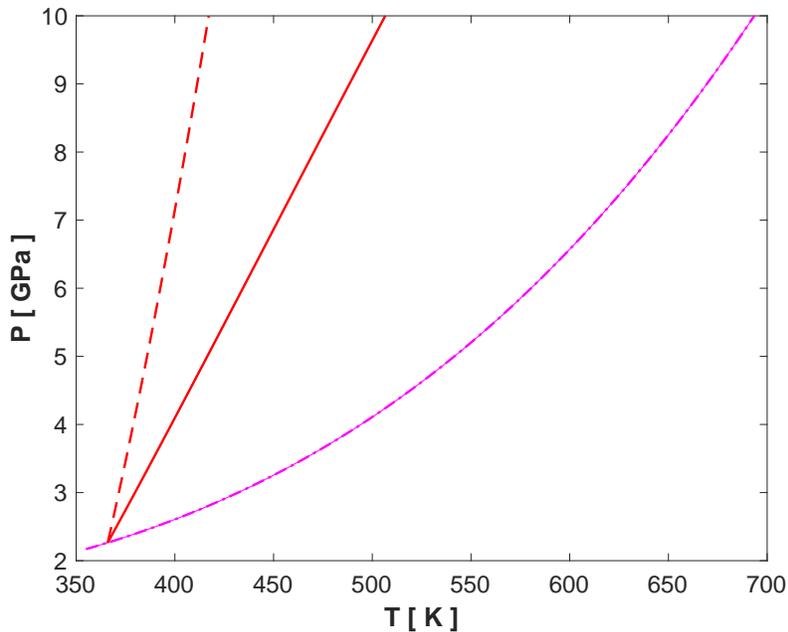}
\caption{\footnotesize{Adiabatic profiles assuming a composition of ice VII (dashed red), and a composition of MH-III (solid red). Dashed-dotted (magenta) is the melting curve of ice VII from \cite{Lin2004}. }}
\label{fig:Adiabats}
\end{figure}

In \cite{Levi2014} we have suggested that the presence of MH-III would shift adiabatic thermal profiles to higher temperatures when compared against adiabats in ice VII. Our new results for the volume thermal expansivity and isobaric heat capacity for MH-III confirm this earlier estimation which was not based on ab initio models for MH-III (see Fig.\ref{fig:Adiabats}). This shift to higher temperatures ought affect the dynamics of ice mantles, and couple between dynamics and composition along the history of CH$_4$ transport and outgassing. However, such an analysis would require a better understanding of the viscosity of high-pressure solid solutions.

The interiors of the icy moons of our solar system reach relatively low pressures. The highest pressure inside Titan is only about $6$\,GPa. Their total mass of volatiles is also likely small in comparison to what is stored inside a ice-rich planet or a super-Moon. Therefore, the high capacity of MH-III for storing volatiles makes it important, even if it is only stabilized over a narrow pressure range ($\sim 0.1$\,GPa). 

Observations beyond our solar system may detect super-Titans down to twice the mass of Mars \citep{Kipping2009,Heller2014}. \cite{Heller2015a,Heller2015b} have shown that super-Titans and super-Ganymedes can form in the accretion disks around super-Jovian planets, noting that hot Jupiters have already been detected. 

Polymorphs of filled ice likely play a major role in the transport and outgassing of volatiles for the case of the more massive super-Titans and water worlds. Filled ices form not just in the H$_2$O-CH$_4$ system (i.e. MH-III) but also in the H$_2$O-CO$_2$ and H$_2$O-N$_2$ systems \citep{Loveday2008}. Thus, separating between biotic and abiotic atmospheric signatures requires a better understanding of these phases and their thermophysical nature. We hope that the community of high-pressure experimentalists and computational material scientists would invest more resources in the study of filled-ices. Such knowledge for the lower pressure clathrate hydrates yielded a general framework for modeling multi-component systems. It is time to do the same for filled-ices if we wish to realistically constrain the uncertainty of biosignatures in water worlds.

\section{SUMMARY}

CH$_4$ is an important biosignature. Therefore, studying its potential internal reservoirs and abiotic origins is needed.  
MH-III (CH$_4$ filled-ice Ih) is a phase that forms in the H$_2$O-CH$_4$ binary mixture above about $1.5$\,GPa, depending on the temperature (see Fig.\ref{fig:PhaseDiagram}). 
In this work we calculate the thermal equation of state and heat capacity of MH-III in the temperature range of $300$\,K-$500$\,K and pressure between $2$\,GPa-$7$\,GPa, using first-principles electronic structure methods. These temperature and pressure regimes are adequate for studying the role of MH-III as a possible reservoir for CH$_4$ in Titan and supr-Titan class objects, speculated to be potentially habitable. This paper focuses on Titan, an endmember in this family of objects. 

We assume for Titan a layered model consisting of a mixed ice-rock layer above a rocky core and underlying a high-pressure ice layer, topped by a subterranean ocean and a crust (see Fig.\ref{fig:TitanSection}). Using observational constraints for Titan we derive the thickness of the different layers as a function of ocean and rock mass density. We show that for a normalized moment of inertia (MOI) of $0.34$ the highest pressure on the outer boundary of the rock core is less than $1.5$\,GPa, hence not allowing for the formation of MH-III. However, for a MOI of $0.33$, and an ocean density of $1.24$\,g\,cm$^{-3}$, MH-III may form above the rocky core if the rock density is higher than $3.05$\,g\,cm$^{-3}$. In this case a layer, about $100$\,km thick, may form above the rocky core that falls inside the thermodynamic stability field of MH-III. Since the mole ratio of H$_2$O:CH$_4$ in MH-III is $2$:$1$ this is potentially a large reservoir for CH$_4$ ($\sim 3000\times$ the surface and atmospheric inventory of CH$_4$ for the case of Titan). 

We use a 1-D thermal evolution model to calculate the heat flux out of Titan's rocky core and into a possible MH-III enriched layer. 
The heat flux out of the core is $\approx 10$\,erg\,cm$^{-2}$\,s$^{-1}$. We show that internal core temperatures are high enough so as to dissociate MH-III out of the core. An exception is a thin ($\approx 10$\,km) layer on the outer core where MH-III may survive to the present between rock grains, likely not holding more than $10^{20}$\,mol of CH$_4$ (assuming a $1$\% porosity). We corroborate, as previously shown by \cite{Choblet2017} and \cite{Kalousova2018}, that melting and melt migration is an important heat transfer mechanism above the rocky core. However, our derived equation of state for MH-III shows it is less dense than liquid water, consequently hindering melt extraction. Melt may need to be further heated to temperatures above the melting temperature in order to become positively buoyant. 

For the case of a mixed MH-III and rock layer, a rock mass fraction higher than $0.3$ would turn melt positively buoyant upon melting. In this case the outward migration of CH$_4$ is likely in the form of a dissolved component within the melt. For low rock mass fractions ($\lessapprox 0.2$) the melt, upon reaching positive buoyancy, should be hot enough to dissociate MH-III along its path, thus transporting CH$_4$ outward more efficiently.  

Water ice VII is a likely polymorph in the interior of super-Titans. Melt at the rock-ice mantle boundary for these larger objects should be positively buoyant upon melting if the mass fraction of ice VII is larger than $0.65$ (mole ratio of H$_2$O:CH$_4$ larger than $5.5$). In this case CH$_4$ likely migrates as a dissolved component within the melt. If, on the other hand, water is less abundant, then positively hot plumes are likely hot enough to dissociate MH-III as they migrate upward, yielding a faster depletion, and potentially outgassing, of internal CH$_4$.

We calculate the heat capacity for MH-III, and show that it likely has a lower Debye temperature compared to pure water ice polymorphs. Together with our derived thermal expansivity coefficient we can calculate adiabats in the interior of super-Titans. We confirm an earlier suggestion made in \cite{Levi2014} that MH-III supports higher internal adiabatic temperature profiles.

\section{ACKNOWLEDGEMENTS}

We wish to thank our referee for important suggestions on how to improve the text.

AL is supported by a grant from the Simons Foundation (SCOL \#290360 to D.S.).
The computations in this paper were run on the Odyssey cluster supported by the FAS Division of Science, Research Computing Group at Harvard University. AL is grateful to the administrative staff for their technical support.   

REC was supported by the European Research Council Advanced Grant ToMCaT and by the Carnegie Institution for Science.
We gratefully acknowledge the Gauss Centre for Supercomputing e.V. (www.gauss-centre.eu) for funding this project in part
by providing computing time on the GCS Supercomputer SuperMUC at Leibniz Supercomputing Centre (LRZ, www.lrz.de).

\section{APPENDIX}

\subsection{Heat Conduction in a Heated Layer}

Heat conduction in 1D with a time-dependent radiogenic heat source may be described with the following differential equation:

\begin{equation}
\kappa\frac{\partial^2T}{\partial x^2}=-\frac{H_{cf}}{C_p}e^{-t/\tau}+\frac{\partial T}{\partial t}
\end{equation}
 where $\kappa$ is the thermal diffusivity coefficient, $T$ is temperature, $H_{cf}$ is the radioactive heat production rate per unit mass at $t=0$, $C_p$ is the isobaric heat capacity, and $\tau$ is an averaged half-life for radioactive decay.
 Assuming a constant and uniform initial temperature of $T_0$, the Laplace transform of the diffusion equation is:
 \begin{equation}
 \kappa\frac{d^2\tilde{T}}{dx^2}-p\tilde{T}+\frac{H_{cf}}{C_p}\frac{1}{p+1/\tau}+T_0=0
\end{equation}
where $p$ is the frequency parameter. The general solution of the last equation is:
\begin{equation}
\tilde{T}(x,p)=Ae^{\sqrt{p/\kappa}x}+Be^{-\sqrt{p/\kappa}x}+\frac{T_0}{p}+\frac{H_{cf}}{C_p}\frac{1}{p(p+1/\tau)}
\end{equation}
We consider a zero flux boundary at $x=0$, and a constant temperature at the boundary $x=d_c$,
\begin{eqnarray}
\frac{\partial T}{\partial x}|_{x=0}=0 \\
T|_{x=d_c}=T_m
\end{eqnarray}
the Laplace transform of the boundary conditions is,

\begin{eqnarray}
\frac{d\tilde{T}}{dx}|_{x=0}=0 \\
\tilde{T}|_{x=d_c}=\frac{T_m}{p}
\end{eqnarray}
giving after a few algebraic steps the solution,

\begin{equation}
\tilde{T}(x,p)=\left(\frac{T_m-T_0}{p}-\frac{H_{cf}}{C_p}\frac{1}{p(p+1/\tau)}\right)\frac{\cosh(qx)}{\cosh(qd_c)}+\frac{T_0}{p}+\frac{H_{cf}}{C_p}\frac{1}{p(p+1/\tau)}
\end{equation}
where $q\equiv\sqrt{p/\kappa}$. The last equation has poles at $p=0$, $p=-1/\tau$, and for $\cosh(qd_c)=0$. Using the inversion theorem we obtain the following solution,
\begin{eqnarray*}
T(x,t)=T_m-\frac{4\left(T_m-T_0\right)}{\pi}\sum^{\infty}_{n=0}\frac{(-1)^n}{2n+1}e^{-\kappa\gamma^2_nt}\cos(\gamma_nx)\\
-\frac{H_{cf}}{C_p}\left\{  \tau e^{-t/\tau}\left[1-\frac{\cos(x/\sqrt{\kappa\tau})}{\cos(d_c/\sqrt{\kappa\tau})}\right]+\sum^{\infty}_{n=0}\frac{4(-1)^{n+1}}{\pi(2n+1)\left[ \frac{1}{\tau}-\kappa\gamma^2_n\right]}e^{-\kappa\gamma^2_nt}\cos(\gamma_nx)  \right\}
\end{eqnarray*}
where,
\begin{equation}
\gamma_n\equiv\frac{(2n+1)\pi}{2d_c}
\end{equation}


\bibliography{amitmemo.bib}

\begin{thebibliography}{}
\expandafter\ifx\csname natexlab\endcsname\relax\def\natexlab#1{#1}\fi

\bibitem[{Baland {et~al.}(2014)Baland, Tobie, Lefèvre, \& Hoolst}]{Baland2014}
Baland, R.-M., Tobie, G., Lefèvre, A., \& Hoolst, T.~V. 2014, Icarus, 237, 29

\bibitem[{Barr {et~al.}(2010)Barr, Citron, \& Canup}]{Barr2010858}
Barr, A.~C., Citron, R.~I., \& Canup, R.~M. 2010, Icarus, 209, 858

\bibitem[{Baumert {et~al.}(2005)Baumert, Gutt, Krisch, Requardt, M\"uller, Tse,
  Klug, \& Press}]{Baumert2005}
Baumert, J., Gutt, C., Krisch, M., {et~al.} 2005, Phys. Rev. B, 72, 054302

\bibitem[{Benoit {et~al.}(1998)Benoit, Marx, \& Parrinello}]{Benoit1998}
Benoit, M., Marx, D., \& Parrinello, M. 1998, Nature, 392, 258

\bibitem[{Berens {et~al.}(1983)Berens, Mackay, White, \& Wilson}]{Berens1983}
Berens, P.~H., Mackay, D. H.~J., White, G.~M., \& Wilson, K.~R. 1983, The
  Journal of Chemical Physics, 79, 2375

\bibitem[{Bethkenhagen {et~al.}(2013)Bethkenhagen, French, \&
  Redmer}]{Bethkenhagen2013}
Bethkenhagen, M., French, M., \& Redmer, R. 2013, The Journal of Chemical
  Physics, 138, 234504

\bibitem[{Bezacier {et~al.}(2014{\natexlab{a}})Bezacier, Journaux, Perrillat,
  Cardon, Hanfland, \& Daniel}]{Bezacier2014}
Bezacier, L., Journaux, B., Perrillat, J.-P., {et~al.} 2014{\natexlab{a}}, The
  Journal of Chemical Physics, 141, doi:http://dx.doi.org/10.1063/1.4894421

\bibitem[{Bezacier {et~al.}(2014{\natexlab{b}})Bezacier, Menn, Grasset,
  Bollengier, Oancea, Mezouar, \& Tobie}]{Bezacier2014b}
Bezacier, L., Menn, E.~L., Grasset, O., {et~al.} 2014{\natexlab{b}}, Physics of
  the Earth and Planetary Interiors, 229, 144

\bibitem[{Brand {et~al.}(2010)Brand, Fortes, Wood, \&
  Vo{\v{c}}adlo}]{Brand2010}
Brand, H. E.~A., Fortes, A.~D., Wood, I.~G., \& Vo{\v{c}}adlo, L. 2010, Physics
  and Chemistry of Minerals, 37, 265

\bibitem[{Bussi {et~al.}(2007)Bussi, Donadio, \& Parrinello}]{Bussi2007}
Bussi, G., Donadio, D., \& Parrinello, M. 2007, The Journal of Chemical
  Physics, 126, 014101

\bibitem[{Castillo-Rogez \& Lunine(2010)}]{Castillo2010}
Castillo-Rogez, J.~C., \& Lunine, J.~I. 2010, Geophysical Research Letters, 37,
  doi:10.1029/2010GL044398

\bibitem[{Chen {et~al.}(2011)Chen, Hsieh, Cahill, Trinkle, \& Li}]{Chen2011}
Chen, B., Hsieh, W.-P., Cahill, D.~G., Trinkle, D.~R., \& Li, J. 2011, Phys.
  Rev. B, 83, 132301

\bibitem[{Choblet {et~al.}(2017)Choblet, Tobie, Sotin, Kalousová, \&
  Grasset}]{Choblet2017}
Choblet, G., Tobie, G., Sotin, C., Kalousová, K., \& Grasset, O. 2017, Icarus,
  285, 252

\bibitem[{Consolmagno~SJ {et~al.}(2006)Consolmagno~SJ, Macke~SJ, Rochette,
  Britt, \& Gattacceca}]{Consolmagn2006}
Consolmagno~SJ, G.~J., Macke~SJ, R.~J., Rochette, P., Britt, D.~T., \&
  Gattacceca, J. 2006, Meteoritics \& Planetary Science, 41, 331

\bibitem[{Davaille \& Jaupart(1993)}]{Davaille1993}
Davaille, A., \& Jaupart, C. 1993, Journal of Fluid Mechanics, 253, 141?166

\bibitem[{Davaille \& Jaupart(1994)}]{Davaille1994}
---. 1994, Journal of Geophysical Research: Solid Earth, 99, 19853

\bibitem[{{Dyadin} {et~al.}(1997){Dyadin}, {Aladko}, \&
  {Larionov}}]{dyadin1997}
{Dyadin}, Y.~A., {Aladko}, E.~Y., \& {Larionov}, E.~G. 1997, Mendeleev Commun.

\bibitem[{{Fei} {et~al.}(1993){Fei}, {Mao}, \& {Hemley}}]{fei1993}
{Fei}, Y., {Mao}, H.-K., \& {Hemley}, R.~J. 1993, J. Chem. Phys., 99, 5369

\bibitem[{{Feistel} \& {Wagner}(2006)}]{feistel06}
{Feistel}, R., \& {Wagner}, W. 2006, J.Phys.Chem.Ref.Data, 35, 1021

\bibitem[{Fortes(2012)}]{Fortes2012c}
Fortes, A. 2012, Planetary and Space Science, 60, 10 , titan Through Time: A
  Workshop on Titan’s Formation, Evolution and Fate

\bibitem[{Frank {et~al.}(2004)Frank, Fei, \& Hu}]{Frank2004}
Frank, M.~R., Fei, Y., \& Hu, J. 2004, Geochimica et Cosmochimica Acta, 68,
  2781

\bibitem[{Frank {et~al.}(2006)Frank, Runge, Scott, Maglio, Olson, Prakapenka,
  \& Shen}]{Frank2006}
Frank, M.~R., Runge, C.~E., Scott, H.~P., {et~al.} 2006, Physics of the Earth
  and Planetary Interiors, 155, 152

\bibitem[{French \& Redmer(2015)}]{French2015}
French, M., \& Redmer, R. 2015, Phys. Rev. B, 91, 014308

\bibitem[{Futera \& English(2018)}]{Futera2018}
Futera, Z., \& English, N.~J. 2018, The Journal of Chemical Physics, 148,
  204505

\bibitem[{Goedecker {et~al.}(1996)Goedecker, Teter, \& Hutter}]{Goedecker1996}
Goedecker, S., Teter, M., \& Hutter, J. 1996, Phys. Rev. B, 54, 1703

\bibitem[{Grasset {et~al.}(2000)Grasset, Sotin, \& Deschamps}]{Grasset2000}
Grasset, O., Sotin, C., \& Deschamps, F. 2000, Planetary and Space Science, 48,
  617

\bibitem[{Grindrod {et~al.}(2008)Grindrod, Fortes, Nimmo, Feltham, Brodholt, \&
  Vočadlo}]{Grindrod2008}
Grindrod, P., Fortes, A., Nimmo, F., {et~al.} 2008, Icarus, 197, 137

\bibitem[{Hartwigsen {et~al.}(1998)Hartwigsen, Goedecker, \&
  Hutter}]{Hartwigsen1998}
Hartwigsen, C., Goedecker, S., \& Hutter, J. 1998, Phys. Rev. B, 58, 3641

\bibitem[{{Heller}(2014)}]{Heller2014}
{Heller}, R. 2014, \apj, 787, 14

\bibitem[{{Heller} \& {Pudritz}(2015{\natexlab{a}})}]{Heller2015a}
{Heller}, R., \& {Pudritz}, R. 2015{\natexlab{a}}, \aap, 578, A19

\bibitem[{{Heller} \& {Pudritz}(2015{\natexlab{b}})}]{Heller2015b}
---. 2015{\natexlab{b}}, \apj, 806, 181

\bibitem[{{Hester} {et~al.}(2007){Hester}, {Huo}, {Ballard}, {Koh}, {Miller},
  \& {Sloan}}]{hester2007}
{Hester}, K.~C., {Huo}, Z., {Ballard}, A.~L., {et~al.} 2007, J. Phys. Chem. B

\bibitem[{Hirai {et~al.}(2006)Hirai, Machida, Kawamura, Yamamoto, \&
  Yagi}]{hirai06}
Hirai, H., Machida, S.-i., Kawamura, T., Yamamoto, Y., \& Yagi, T. 2006,
  American Mineralogist, 91, 826

\bibitem[{{Hirai} {et~al.}(2003){Hirai}, {Tanaka}, {Kawamura}, {Yamamoto}, \&
  {Yagi}}]{hirai2003}
{Hirai}, H., {Tanaka}, T., {Kawamura}, T., {Yamamoto}, Y., \& {Yagi}, T. 2003,
  Phys. Rev. B, 68, 172102

\bibitem[{{Hirai} {et~al.}(2001){Hirai}, {Uchihara}, {Fujihisa}, M., {Katoh},
  {Aoki}, {Nagashima}, {Yamamoto}, \& {Yagi}}]{hirai01}
{Hirai}, H., {Uchihara}, Y., {Fujihisa}, K., {et~al.} 2001, J. Chem. Phys.,
  115, 7066

\bibitem[{ichi Machida {et~al.}(2006)ichi Machida, Hirai, Kawamura, Yamamoto,
  \& Yagi}]{Machida2006}
ichi Machida, S., Hirai, H., Kawamura, T., Yamamoto, Y., \& Yagi, T. 2006,
  Physics of the Earth and Planetary Interiors, 155, 170

\bibitem[{Iess {et~al.}(2010)Iess, Rappaport, Jacobson, Racioppa, Stevenson,
  Tortora, Armstrong, \& Asmar}]{Iess2010}
Iess, L., Rappaport, N.~J., Jacobson, R.~A., {et~al.} 2010, Science, 327, 1367

\bibitem[{{Iess} {et~al.}(2012){Iess}, {Jacobson}, {Ducci}, {Stevenson},
  {Lunine}, {Armstrong}, {Asmar}, {Racioppa}, {Rappaport}, \&
  {Tortora}}]{Iess2012}
{Iess}, L., {Jacobson}, R.~A., {Ducci}, M., {et~al.} 2012, Science, 337, 457

\bibitem[{Jelen {et~al.}(2016)Jelen, Giovannelli, \& Falkowski}]{Jelen2016}
Jelen, B.~I., Giovannelli, D., \& Falkowski, P.~G. 2016, Annual Review of
  Microbiology, 70, 45, pMID: 27297124

\bibitem[{Journaux {et~al.}(2017)Journaux, Daniel, Petitgirard, Cardon,
  Perrillat, Caracas, \& Mezouar}]{Journaux2017}
Journaux, B., Daniel, I., Petitgirard, S., {et~al.} 2017, Earth and Planetary
  Science Letters, 463, 36

\bibitem[{Kadobayashi {et~al.}(2018)Kadobayashi, Hirai, Ohfuji, Ohtake, \&
  Yamamoto}]{Kadobayashi2018}
Kadobayashi, H., Hirai, H., Ohfuji, H., Ohtake, M., \& Yamamoto, Y. 2018, The
  Journal of Chemical Physics, 148, 164503

\bibitem[{Kalousov\'{a} {et~al.}(2018)Kalousov\'{a}, Sotin, Choblet, Tobie, \&
  Grasset}]{Kalousova2018}
Kalousov\'{a}, K., Sotin, C., Choblet, G., Tobie, G., \& Grasset, O. 2018,
  Icarus, 299, 133

\bibitem[{Kaltenegger {et~al.}(2010)Kaltenegger, Selsis, Fridlund, Lammer,
  Beichman, Danchi, Eiroa, Henning, Herbst, Léger, Liseau, Lunine, Paresce,
  Penny, Quirrenbach, Röttgering, Schneider, Stam, Tinetti, \&
  White}]{Kaltenegger2010}
Kaltenegger, L., Selsis, F., Fridlund, M., {et~al.} 2010, Astrobiology, 10, 89,
  pMID: 20307185

\bibitem[{Kipping {et~al.}(2009)Kipping, Fossey, \& Campanella}]{Kipping2009}
Kipping, D.~M., Fossey, S.~J., \& Campanella, G. 2009, Monthly Notices of the
  Royal Astronomical Society, 400, 398

\bibitem[{Kirk \& Stevenson(1987)}]{Kirk1987}
Kirk, R., \& Stevenson, D. 1987, Icarus, 69, 91

\bibitem[{Krack(2005)}]{Krack2005}
Krack, M. 2005, Theoretical Chemistry Accounts, 114, 145

\bibitem[{Kurnosov {et~al.}(2006)Kurnosov, Dubrovinsky, Kuznetsov, \&
  Dmitriev}]{Kurnosov2006}
Kurnosov, A., Dubrovinsky, L., Kuznetsov, A., \& Dmitriev, V. 2006, zeitschrift
  fur Naturforschung B, 61, 1573

\bibitem[{{Landau} \& {Lifshitz}(2007)}]{lanlif5}
{Landau}, L.~D., \& {Lifshitz}, E.~M. 2007, {Statistical physics Part 1}, 3rd
  edn., Vol.~5 (Butterworth-Heinemann)

\bibitem[{Lee {et~al.}(2010)Lee, Murray, Kong, Lundqvist, \&
  Langreth}]{Kyuho2010}
Lee, K., Murray, E.~D., Kong, L., Lundqvist, B.~I., \& Langreth, D.~C. 2010,
  Phys. Rev. B, 82, 081101

\bibitem[{Lejaeghere {et~al.}(2014)Lejaeghere, Speybroeck, Oost, \&
  Cottenier}]{Lejaeghere2014}
Lejaeghere, K., Speybroeck, V.~V., Oost, G.~V., \& Cottenier, S. 2014, Critical
  Reviews in Solid State and Materials Sciences, 39, 1

\bibitem[{{Levi} {et~al.}(2013){Levi}, {Sasselov}, \& {Podolak}}]{Levi2013}
{Levi}, A., {Sasselov}, D., \& {Podolak}, M. 2013, The Astrophysical Journal,
  769, 29

\bibitem[{Levi {et~al.}(2014)Levi, Sasselov, \& Podolak}]{Levi2014}
Levi, A., Sasselov, D., \& Podolak, M. 2014, The Astrophysical Journal, 792,
  125

\bibitem[{{Lin} {et~al.}(2004){Lin}, {Militzer}, {Struzhkin}, {Gregoryanz},
  {Hemley}, \& {Mao}}]{Lin2004}
{Lin}, J.-F., {Militzer}, B., {Struzhkin}, V.~V., {et~al.} 2004, J. Chem.
  Phys., 121, 8423

\bibitem[{Lorenz {et~al.}(2008)Lorenz, Mitchell, Kirk, Hayes, Aharonson,
  Zebker, Paillou, Radebaugh, Lunine, Janssen, Wall, Lopes, Stiles, Ostro,
  Mitri, \& Stofan}]{Lorenz2008}
Lorenz, R.~D., Mitchell, K.~L., Kirk, R.~L., {et~al.} 2008, Geophysical
  Research Letters, 35,
  https://agupubs.onlinelibrary.wiley.com/doi/pdf/10.1029/2007GL032118

\bibitem[{Loveday \& Nelmes(2008)}]{Loveday2008}
Loveday, J.~S., \& Nelmes, R.~J. 2008, Phys. Chem. Chem. Phys., 10, 937

\bibitem[{{Loveday} {et~al.}(2001{\natexlab{a}}){Loveday}, {Nelmes}, {Guthrie},
  A., {Allan}, {Klug}, {Tse}, \& {Handa}}]{lovedaynat01}
{Loveday}, J.~S., {Nelmes}, R.~J., {Guthrie}, M., {et~al.} 2001{\natexlab{a}},
  Let. Nat., 410, 661

\bibitem[{{Loveday} {et~al.}(2001{\natexlab{b}}){Loveday}, {Nelmes}, {Guthrie},
  D., \& {Tse}}]{loveday01}
{Loveday}, J.~S., {Nelmes}, R.~J., {Guthrie}, M., D., K.~D., \& {Tse}, J.~S.
  2001{\natexlab{b}}, Phys. Rev. Let., 87, 215501(1)

\bibitem[{{Lunine} {et~al.}(2010){Lunine}, {Choukroun}, {Stevenson}, \&
  {Tobie}}]{lunine10}
{Lunine}, J., {Choukroun}, M., {Stevenson}, D., \& {Tobie}, G. 2010, {The
  Origin and Evolution of Titan}, ed. {Brown, R.~H., Lebreton, J.-P., \& Waite,
  J.~H.}, 35--+

\bibitem[{Lunine(2009)}]{Lunine2009}
Lunine, J.~I. 2009, Proceedings of the American Philosophical Society, 153, 403

\bibitem[{Lunine \& Stevenson(1987)}]{Lunine198761}
Lunine, J.~I., \& Stevenson, D.~J. 1987, Icarus, 70, 61

\bibitem[{Macke {et~al.}(2011)Macke, Consolmagno, \& Britt}]{Macke2011}
Macke, R.~J., Consolmagno, G.~J., \& Britt, D.~T. 2011, Meteoritics \&
  Planetary Science, 46, 1842

\bibitem[{Marques {et~al.}(2012)Marques, Oliveira, \& Burnus}]{Marques2012}
Marques, M., Oliveira, M., \& Burnus, T. 2012, Computer Physics Communications,
  183, 2272

\bibitem[{McKay(2014)}]{McKay2014}
McKay, C.~P. 2014, Proceedings of the National Academy of Sciences, 111, 12628

\bibitem[{{McKay}(2016)}]{McKay2016}
{McKay}, C.~P. 2016, Life, 6, 8

\bibitem[{Mitri {et~al.}(2014)Mitri, Meriggiola, Hayes, Lefevre, Tobie, Genova,
  Lunine, \& Zebker}]{Mitri2014}
Mitri, G., Meriggiola, R., Hayes, A., {et~al.} 2014, Icarus, 236, 169

\bibitem[{Monteux {et~al.}(2014)Monteux, Tobie, Choblet, \&
  Feuvre}]{Monteux2014377}
Monteux, J., Tobie, G., Choblet, G., \& Feuvre, M.~L. 2014, Icarus, 237, 377

\bibitem[{Myint {et~al.}(2017)Myint, Benedict, \& Belof}]{Myint2017}
Myint, P.~C., Benedict, L.~X., \& Belof, J.~L. 2017, The Journal of Chemical
  Physics, 147, 084505

\bibitem[{Niemann {et~al.}(2010)Niemann, Atreya, Demick, Gautier, Haberman,
  Harpold, Kasprzak, Lunine, Owen, \& Raulin}]{Niemann2010}
Niemann, H.~B., Atreya, S.~K., Demick, J.~E., {et~al.} 2010, Journal of
  Geophysical Research: Planets, 115,
  https://agupubs.onlinelibrary.wiley.com/doi/pdf/10.1029/2010JE003659

\bibitem[{Nimmo \& Bills(2010)}]{Nimmo2010}
Nimmo, F., \& Bills, B. 2010, Icarus, 208, 896

\bibitem[{Qi \& Reed(2012)}]{Tingting2012}
Qi, T., \& Reed, E.~J. 2012, The Journal of Physical Chemistry A, 116, 10451,
  pMID: 23013329

\bibitem[{Reese {et~al.}(2005)Reese, Solomatov, \& Baumgardner}]{Reese2005}
Reese, C., Solomatov, V., \& Baumgardner, J. 2005, Physics of the Earth and
  Planetary Interiors, 149, 361

\bibitem[{{Schubert} {et~al.}(2001){Schubert}, {Turcotte}, \&
  {Olson}}]{Schubert2001}
{Schubert}, G., {Turcotte}, D.~L., \& {Olson}, P. 2001, {Mantle Convection in
  the Earth and Planets}

\bibitem[{{Shin} {et~al.}(2012){Shin}, {Kumar}, {Udachin}, {Alavi}, \&
  {Ripmeester}}]{Shin2012}
{Shin}, K., {Kumar}, R., {Udachin}, K.~A., {Alavi}, S., \& {Ripmeester}, J.~A.
  2012, Proceedings of the National Academy of Science, 109, 14785

\bibitem[{{Shin} {et~al.}(2013){Shin}, {Udachin}, {Moudrakovski}, {Leek},
  {Alavi}, {Ratcliffe}, \& {Ripmeester}}]{Shin2013}
{Shin}, K., {Udachin}, K.~A., {Moudrakovski}, I.~L., {et~al.} 2013, Proceedings
  of the National Academy of Science, 110, 8437

\bibitem[{Sohl {et~al.}(2003)Sohl, Hussmann, Schwentker, Spohn, \&
  Lorenz}]{Sohl2003}
Sohl, F., Hussmann, H., Schwentker, B., Spohn, T., \& Lorenz, R.~D. 2003,
  Journal of Geophysical Research: Planets, 108, n/a, 5130

\bibitem[{{Solomatov}(1995)}]{Solomatov1995}
{Solomatov}, V.~S. 1995, Physics of Fluids, 7, 266

\bibitem[{Tchijov(2004)}]{Tchijov2004}
Tchijov, V. 2004, Journal of Physics and Chemistry of Solids, 65, 851

\bibitem[{{Thiriet} {et~al.}(2019){Thiriet}, {Breuer}, {Michaut}, \&
  {Plesa}}]{Thiriet2019}
{Thiriet}, M., {Breuer}, D., {Michaut}, C., \& {Plesa}, A.-C. 2019, Physics of
  the Earth and Planetary Interiors, 286, 138

\bibitem[{Tobie {et~al.}(2012)Tobie, Gautier, \& Hersant}]{Tobie2012}
Tobie, G., Gautier, D., \& Hersant, F. 2012, The Astrophysical Journal, 752,
  125

\bibitem[{VandeVondele \& Hutter(2007)}]{VandeVondele2007}
VandeVondele, J., \& Hutter, J. 2007, The Journal of Chemical Physics, 127,
  114105

\bibitem[{VandeVondele {et~al.}(2005)VandeVondele, Krack, Mohamed, Parrinello,
  Chassaing, \& Hutter}]{VandVondele2005}
VandeVondele, J., Krack, M., Mohamed, F., {et~al.} 2005, Computer Physics
  Communications, 167, 103

\bibitem[{Vosko {et~al.}(1980)Vosko, Wilk, \& Nusair}]{VWN}
Vosko, S.~H., Wilk, L., \& Nusair, M. 1980, Canadian Journal of Physics, 58,
  1200

\bibitem[{{Wagner} \& {Pruss}(2002)}]{wagner02}
{Wagner}, W., \& {Pruss}, A. 2002, J.Phys.Chem.Ref.Data, 31, 387

\bibitem[{Yomogida \& Matsui(1983)}]{Yomogida1983}
Yomogida, K., \& Matsui, T. 1983, Journal of Geophysical Research: Solid Earth,
  88, 9513

\end{thebibliography}

\end{document}